\documentclass[twocolumn,tighten,twocolappendix]{aastex63}
\pdfoutput=1 %for arXiv submission
\usepackage{amsmath,amstext}
\usepackage[T1]{fontenc}
\usepackage{ae,aecompl}
\usepackage[utf8]{inputenc}
\usepackage{apjfonts} 
\usepackage[figure,figure*]{hypcap}
\usepackage{enumitem}
\usepackage{bm}
\usepackage{graphicx}
\usepackage{lineno}
\usepackage{tikz}

\definecolor{wdm_3_delta_08}{HTML}{c994c7}
\definecolor{wdm_4_delta_08}{HTML}{df65b0}
\definecolor{wdm_5_delta_08}{HTML}{e7298a}
\definecolor{wdm_6_delta_08}{HTML}{ce1256}
\definecolor{wdm_6_5_delta_08}{HTML}{980043}
\definecolor{wdm_10_delta_08}{HTML}{67001f}

\definecolor{wdm_3_delta_06}{HTML}{a1d99b}
\definecolor{wdm_4_delta_06}{HTML}{74c476}
\definecolor{wdm_5_delta_06}{HTML}{41ab5d}
\definecolor{wdm_6_delta_06}{HTML}{238b45}
\definecolor{wdm_6_5_delta_06}{HTML}{006d2c}
\definecolor{wdm_10_delta_06}{HTML}{00441b}

\definecolor{wdm_3_delta_04}{HTML}{9ecae1}
\definecolor{wdm_4_delta_04}{HTML}{6baed6}
\definecolor{wdm_5_delta_04}{HTML}{4292c6}
\definecolor{wdm_6_delta_04}{HTML}{2171b5}
\definecolor{wdm_6_5_delta_04}{HTML}{08519c}
\definecolor{wdm_10_delta_04}{HTML}{08306b}

\definecolor{wdm_3_delta_02}{HTML}{bcbddc}
\definecolor{wdm_4_delta_02}{HTML}{9e9ac8}
\definecolor{wdm_5_delta_02}{HTML}{807dba}
\definecolor{wdm_6_delta_02}{HTML}{6a51a3}
\definecolor{wdm_6_5_delta_02}{HTML}{54278f}
\definecolor{wdm_10_delta_02}{HTML}{3f007d}

\input{hyperlink-year-only-natbib-patch}

\newcommand*{\http}[1]{\href{http://#1}{#1}}
\newcommand*{\https}[1]{\href{https://#1}{#1}}

\shorttitle{COZMIC. II.}
\shortauthors{An et al.}

\begin{document}

\title{COZMIC. II. Cosmological Zoom-in Simulations with Fractional non-CDM Initial Conditions}

\author[0000-0001-9543-5012]{Rui An}
\affiliation{Department of Physics $\&$ Astronomy, University of Southern California, Los Angeles, CA 90007, USA}

\author[0000-0002-1182-3825]{Ethan O.~Nadler}
\affiliation{Department of Astronomy \& Astrophysics, University of California, San Diego, La Jolla, CA 92093, USA}
\affiliation{Carnegie Observatories, 813 Santa Barbara Street, Pasadena, CA 91101, USA}
\affiliation{Department of Physics $\&$ Astronomy, University of Southern California, Los Angeles, CA 90007, USA}

\author[0000-0001-5501-6008]{Andrew Benson}
\affiliation{Carnegie Observatories, 813 Santa Barbara Street, Pasadena, CA 91101, USA}

\author[0000-0002-3589-8637]{Vera Gluscevic}
\affiliation{Department of Physics $\&$ Astronomy, University of Southern California, Los Angeles, CA 90007, USA}

\correspondingauthor{Ethan O.~Nadler}
\email{enadler@ucsd.edu}

\label{firstpage}

\begin{abstract}
We present $24$ cosmological dark matter (DM)-only zoom-in simulations of a Milky Way analog with initial conditions appropriate for scenarios where non-cold dark matter (NCDM) is a fraction of the total DM abundance (f-NCDM models) as the second installment of the COZMIC suite. We initialize our simulations using transfer functions, $\mathcal{T}_{\mathrm{f-NCDM}}(k)\equiv\sqrt{P_{\mathrm{f-NCDM}}(k)/P_{\mathrm{CDM}}(k)}$ (where $P(k)$ is the linear matter power spectrum), with an initial suppression similar to thermal-relic warm dark matter (WDM) followed by a constant-amplitude plateau. We simulate suppression wavenumbers $[22.8,~ 32.1,~ 41.8,~ 52.0,~ 57.1,~ 95.3]~\mathrm{Mpc}^{-1}$, corresponding to thermal-relic WDM masses $m_{\mathrm{WDM}}\in [3,~ 4,~ 5,~ 6,~ 6.5,~ 10]~\mathrm{keV}$, and plateau amplitudes $\delta\in [0.2,~ 0.4,~ 0.6,~ 0.8]$. We model the subhalo mass function in terms of the suppression wavenumber and $\delta$. Integrating these models into a forward model of the MW satellite galaxy population yields new limits on f-NCDM scenarios, with suppression wavenumbers greater than $46$ and $ 40~\mathrm{Mpc}^{-1}$ for $\delta=0.2$, $0.4$, respectively, at $95\%$ confidence. The current data do not constrain $\delta>0.4$. We map these limits to scenarios where a fraction $f_{\mathrm{WDM}}$ of DM behaves as a thermal relic, which yields the following bounds on cosmologies with a mixture of WDM and CDM: $m_{\mathrm{WDM}}>3.6,~ 4.1,~ 4.6,~ 4.9,~ 5.4~\mathrm{keV}$ for $f_{\mathrm{WDM}}=0.5,~ 0.6,~ 0.7,~ 0.8,~ 0.9$, respectively, at 95\% confidence. The current data do not constrain WDM fractions $f_{\mathrm{WDM}}<0.5$. Our results affirm that low-mass halo abundances are sensitive to partial suppression in $P(k)$, indicating the possibility of using galactic substructure to reconstruct $P(k)$ on small scales.
\end{abstract}

\keywords{\href{http://astrothesaurus.org/uat/1787}{Warm dark matter (1787)}; 
\href{http://astrothesaurus.org/uat/574}{Galaxy abundances (574)};
\href{http://astrothesaurus.org/uat/1049}{Milky Way dark matter halo (1049)}; 
\href{http://astrothesaurus.org/uat/1083}{$N$-body simulations (1083)};
\href{http://astrothesaurus.org/uat/1880}{Galaxy dark matter halos (1880)}}

\section{Introduction}\label{sec:intro}

Recent observations of small-scale cosmic structure probe dark matter (DM) microphysics beyond collisionless cold dark matter (CDM) with unprecedented sensitivity (e.g., see \citealt{Bullock170704256,Gluscevic190305140,Banerjee220307049,Bechtol220307354} for reviews). In the most commonly considered beyond-CDM scenarios the entire DM relic density is a non-CDM species. In such scenarios, small-scale linear density perturbations are significantly suppressed, entirely suppressing the linear matter power spectrum $P(k)$ below a certain scale (i.e., above a certain wavenumber $k$).

However, in many well-motivated DM models, only a fraction of DM behaves differently from CDM. We refer to these as fractional non-cold dark matter (f-NCDM) scenarios. They typically feature only a partial suppression of $P(k)$ on small scales. The f-NCDM component can suppress small-scale power through a range of physical mechanisms, including free streaming, diffusion damping, and wave interference. Early studies considered a component of free-streaming hot dark matter (HDM; e.g., Standard Model neutrinos; \citealt{Klypin9305011,Liddle9304017}). More recently, fractional warm dark matter (f-WDM) scenarios have gained popularity. For example, sterile neutrino DM models often feature multiple generations of neutrinos with different masses and thus with different free-streaming scales (e.g., see \citealt{Kusenko09062968,Abazajian170501837}). Meanwhile, the string theory axiverse predicts a wide mass spectrum of ultralight fuzzy dark matter (FDM) scalar fields, only some of which do not cluster on small scales \citep{Arvanitaki09054720,Rogers230108361,Winch240411071}. Cosmic tensions on larger scales have also motivated scenarios where a fraction of the DM is collisional; these models feature $P(k)$ suppression due to an f-NCDM component (e.g., \citealt{Brinckmann221213264,He230108260,Rubira220903974}). Thus, it is important to robustly model small-scale structure in f-NCDM scenarios.

To understand how f-NCDM models impact $P(k)$, it is instructive to consider the case where a fraction of the DM free streams, as in f-WDM models. Linear Boltzmann calculation for a mixture of CDM and WDM (or CDM and HDM) predicts small-scale $P(k)$ suppression \citep{Boyanovsky07110470}. This suppression can be significant even when a modest fraction of the DM free streams, since both free-streaming itself and its backreaction on the CDM component suppress the growth of density perturbations. For example, $P(k)/P_{\mathrm{CDM}}(k)$ plateaus to $\approx (1-14 f_{\mathrm{WDM}})$ on small scales when CDM is mixed with a subdominant WDM species, with WDM fraction $f_{\mathrm{WDM}}\ll 1$ \citep{Boyarsky08120010}; a similar mechanism has been used to constrain neutrino masses using large-scale structure. On smaller, nonlinear scales, simulations have been performed for thermal-relic f-WDM scenarios (e.g., \citealt{Anderhalden12122967,Harada14121592,Parimbelli210604588}). These simulations reveal substantial (sub)halo mass function suppression relative to CDM, along with a reduction in the concentrations of low-mass (sub)halos \citep{Maccio12022858} and the local DM density \citep{Anderhalden12063788}.

To date, small-scale structure constraints on f-NCDM scenarios have focused largely on f-WDM, which has been constrained using the Ly$\alpha$ forest \citep{Boyarsky08120010}, Milky Way (MW) satellite galaxy abundances \citep{Anderhalden12122967}, strong gravitational lensing \citep{Kamada160401489}, and combinations thereof (e.g., \citealt{Diamanti170103128}). These studies demonstrate that small-scale structure is sensitive to $P(k)$ suppression even in the absence of a complete cutoff, although they generally lose sensitivity below a certain WDM fraction. In this context, there are several key considerations that motivate our work. First, a relatively small number of simulations have been performed in f-NCDM scenarios, and most of these lack the resolution to inform recent small-scale structure measurements, which currently probe (sub)halos down to masses of $\approx 10^8~M_{\mathrm{\odot}}$ \citep{Drlica-Wagner190201055,Bechtol220307354}. Second, most f-NCDM studies consider f-WDM, despite the wide range of DM microphysics that can give rise to f-NCDM phenomenology, as discussed above. A high-resolution (HR) simulation effort that informs a variety of f-NCDM scenarios is therefore timely.

To address these points, we present $24$ HR cosmological DM-only zoom-in simulations of an MW analog in f-NCDM scenarios, as the second installment of the \textbf{CO}smological \textbf{Z}oo\textbf{M}-in simulations with \textbf{I}nitial \textbf{C}onditions beyond CDM (COZMIC) suite, following the first installment in \defcitealias{Nadler241003635}{Paper~I} Nadler et al.\ (\citeyear{Nadler241003635}, hereafter \citetalias{Nadler241003635}). Our host halo is drawn from the Milky Way-est suite \citep{Buch240408043} and features a realistic LMC analog subhalo and accretion history; as a result, our predictions are directly relevant for MW subhalo and satellite population predictions. Furthermore, our simulations resolve subhalos down to $\approx 10^8~M_{\mathrm{\odot}}$, and therefore probe the comoving scales that drive current small-scale DM constraints. We will leverage this resolution and sample size to derive a fitting function for the subhalo mass function (SHMF) in f-NCDM scenarios. By integrating this fit into a forward model of the MW satellite galaxy population observed by the Dark Energy Survey (DES) and Pan-STARRS1, as compiled in \cite{Drlica-Wagner191203302}, we derive new constraints on f-NCDM transfer functions. We also translate these constraints to f-WDM. These results demonstrate that MW satellite abundances are sensitive to a wide range of f-NCDM scenarios.

This paper is organized as follows. In Section~\ref{sec:model}, we describe our f-NCDM model; in Section~\ref{sec:simulations}, we present our simulation pipeline; in Section~\ref{sec:basic}, we measure SHMFs; in Section~\ref{sec:shmf}, we fit f-NCDM SHMF suppression. We derive bounds on f-NCDM transfer functions and f-WDM in Section~\ref{sec:limits}; we discuss our results in Section~\ref{sec:discussion} and conclude in Section~\ref{sec:conclusions}. Appendices present additional simulation results (Appendix~\ref{sec:additional}) and convergence tests (Appendix~\ref{sec:convergence}).

We use cosmological parameters $h = 0.7$, $\Omega_{\rm m} = 0.286$, $\Omega_b = 0.049$, $\Omega_{\Lambda} = 0.714$, $\sigma_8 = 0.82$, and $n_s=0.96$ \citep{Hinshaw_2013}. Halo masses are defined via the \cite{Bryan_1998} virial overdensity, which corresponds to $\Delta_{\mathrm{vir}}\approx 99\times \rho_{\mathrm{crit}}$ in our cosmology, where $\rho_{\mathrm{crit}}$ is the critical density of the universe at $z=0$. In this work, we study only those subhalos within the virial radius of the MW host. Throughout, ``log'' always refers to the base-$10$ logarithm.

%%%%%%%%%%%%%%%%%%%%%%
%%%%%%%%%%%%%%%%%%%%%%

\section{f-NCDM Model}
\label{sec:model}

We describe the suppression of the linear matter power spectrum $P(k)$ through the transfer function
\begin{equation}
    \mathcal{T}(k) \equiv \sqrt{P(k)/P_\mathrm{CDM}(k)}
\end{equation}
where $P_\mathrm{CDM}(k)$ is the CDM power spectrum and $k$ is the comoving wavenumber. We parameterize f-NCDM transfer functions by a characteristic scale at which $P(k)$ is suppressed $\alpha$, and the height of a constant-amplitude $\mathcal{T}(k)$ plateau $\delta$. Specifically, we use \citep{Hooper2022}
\begin{equation}\label{eq.Twdm}
   \mathcal{T}_{\mathrm{f-NCDM}}(k,\alpha,\delta) = (1-\delta)\left[1+(\alpha k)^{2\nu}\right]^{-5/\nu}+\delta.
\end{equation}

\begin{figure*}[t!]
\centering
\includegraphics[width=0.8\textwidth]{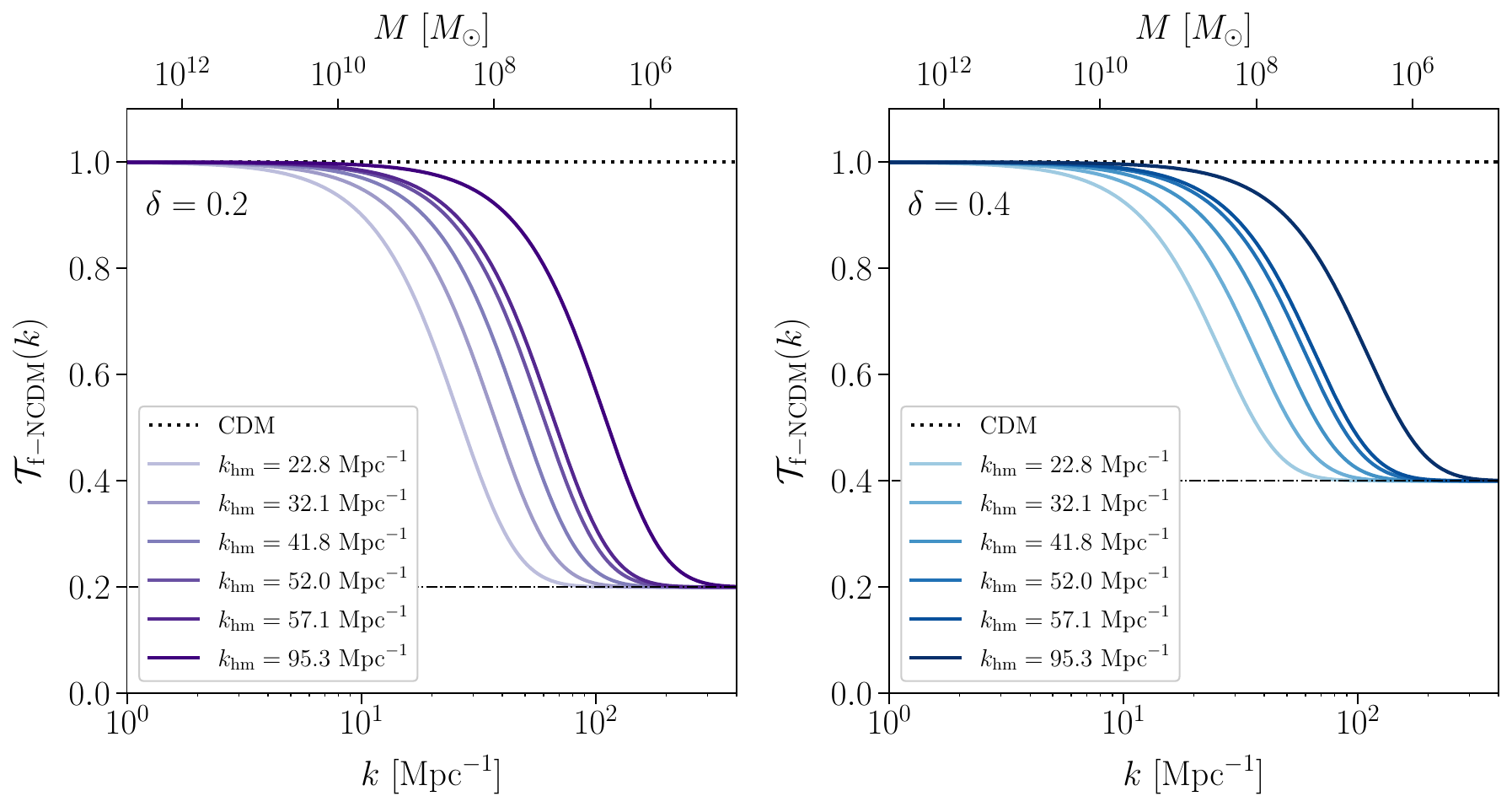}\\
\includegraphics[width=0.8\textwidth]{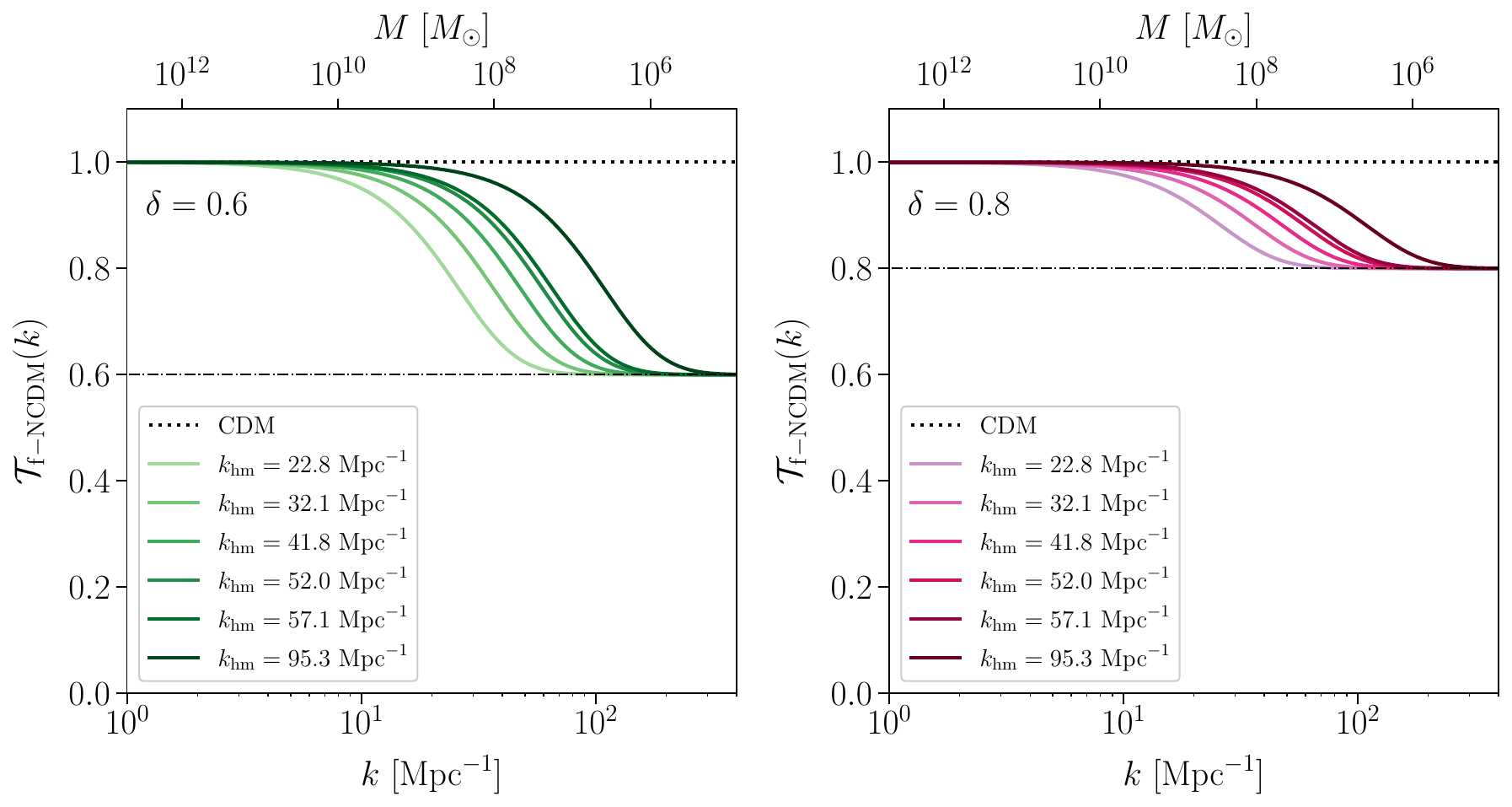}
    \caption{Transfer functions for f-NCDM models simulated in this work (solid colored lines) and for CDM (dotted black lines). Models are ordered from large suppression scale (smallest $k_{\mathrm{hm}}$; light colors) to small suppression scale (largest $k_{\mathrm{hm}}$; dark colors). Dotted-dashed black lines show the height of each transfer function plateau, $\delta$. Top ticks show halo masses associated with wavenumbers in linear theory (Equation~\ref{eq:m_k}).}
    \label{fig:transfers}
\end{figure*}

We assume that the shape of the initial suppression in $P(k)$ follows that in $100\%$ thermal-relic WDM models, for which the transfer function $\mathcal{T}_\mathrm{WDM}(k) = \sqrt{P_\mathrm{WDM}(k)/P_\mathrm{CDM}(k)}$ can be parameterized as \citep{Viel0501562}
\begin{equation}
   \mathcal{T}_{\mathrm{WDM}}(k,m_{\mathrm{WDM}}) = \left[1+(\alpha k)^{2\nu}\right]^{-5/\nu}.\label{eq:t_wdm}
\end{equation}
Here, $\alpha(m_{\mathrm{WDM}})$ parameterizes the suppression scale and $\nu=1.049$ is fixed. We use the fitting function for  $\alpha(m_{\mathrm{WDM}})$ from \cite{Vogel221010753},
\begin{equation}
    \alpha(m_{\mathrm{WDM}})=a \left(\frac{m_{\mathrm{WDM}}}{1~\mathrm{keV}}\right)^b \left(\frac{\omega_{\mathrm{WDM}}}{0.12}\right)^{\eta}\left(\frac{h}{0.6736}\right)^{\theta}h^{-1}~\mathrm{Mpc},\label{eq:alpha_wdm}
\end{equation}
where $a=0.0437$, $b=-1.188$, $\theta=2.012$, $\eta=0.2463$, and $\omega_{\mathrm{WDM}}\equiv \Omega_{\mathrm{WDM}}h^2$. We always evaluate Equation~\ref{eq:alpha_wdm} for a $100\%$ WDM cosmology to determine the shape of the suppression, i.e., we set $\omega_{\mathrm{WDM}}=\omega_{\mathrm{m}}$.

We relate the suppression scale in $P(k)$ to the half-mode wavenumber $k_{\mathrm{hm}}$, defined by $\mathcal{T}(k_{\mathrm{hm}})\equiv 0.5$. Combining this definition with Equations~\ref{eq:t_wdm} and \ref{eq:alpha_wdm} yields
\begin{equation}
    k_{\mathrm{hm}}(m_{\mathrm{WDM}})= 22.7\times \left(\frac{m_{\mathrm{WDM}}}{3~\mathrm{keV}}\right)^{1.188}~\mathrm{Mpc}^{-1}.
\end{equation}
It is useful to relate wavenumbers to halo mass scales; in linear theory, this relation is \citep{Nadler190410000}
\begin{equation}
    M(k) \equiv \frac{4\pi}{3}\Omega_m\bar{\rho}\left(\frac{\pi}{k}\right)^3=5.1\times 10^9\times\left(\frac{10~\mathrm{Mpc}^{-1}}{k}\right)^3~M_{\mathrm{\odot}}.\label{eq:m_k}
\end{equation}
From Equations~\ref{eq:alpha_wdm} and \ref{eq:m_k}, we derive the half-mode mass,
\begin{equation}
    M_{\mathrm{hm}}(m_{\mathrm{WDM}}) = 4.3\times 10^8\times\left(\frac{m_{\mathrm{WDM}}}{3~\mathrm{keV}}\right)^{-3.564}~M_{\mathrm{\odot}}.\label{eq:Mhm_mwdm}
\end{equation}

We simulate transfer functions with values of $k_{\mathrm{hm}}$ that correspond to specific values of  $m_{\mathrm{WDM}}$. Thus, our transfer functions can be written as
\begin{equation}
   \mathcal{T}_{\mathrm{f-NCDM}}(k,k_{\mathrm{hm}},\delta) = (1-\delta)\times \mathcal{T}_{\mathrm{WDM}}(k,m_{\mathrm{WDM}}(k_{\mathrm{hm}}))+\delta.\label{eq:t_fncdm_final}
\end{equation}
Note that the NCDM fraction is $f_{\mathrm{NCDM}}=1-\delta$. Thus, $\delta=0$ corresponds to $100\%$ WDM, which was simulated in \citetalias{Nadler241003635}, and $\delta=1$ corresponds to CDM, which was simulated in \cite{Buch240408043}. We note that in models with $\delta > 0.5$ the transfer function never drops to one-half its CDM value, and therefore, the characteristic suppression scale $k_{\mathrm{hm}}$ does not have a direct physical interpretation; however, we still retain this parameterization for fractional models, for consistency.

Transfer functions for f-WDM have a similar suppression shape compared to our f-NCDM transfer functions, by construction, and then slowly decrease in power on small scales rather than reaching a constant-amplitude plateau. This is due to the backreaction of the non-clustering matter on the remaining, cold component: a fraction of the DM streams out of gravitational potential wells, which reduces the growth rate of CDM density perturbations (e.g., see \citealt{Boyarsky08120010}). Nonetheless, we show that our f-NCDM transfer functions provide a good approximation to f-WDM transfer functions when mapping our f-NCDM constraints to these scenarios in Section~\ref{sec:limits}, for the observables we consider in this study. Meanwhile, models where a fraction of DM interacts with baryons generally have transfer functions with a slightly different suppression shape but do reach constant-amplitude plateaus (e.g., \citealt{He230108260}). We leave a mapping of our bounds to such scenarios for future work.

%%%%%%%%%%%%%%%%%%%%%%

\section{Simulation Pipeline}
\label{sec:simulations}

We now describe our f-NCDM simulation pipeline, including initial condition (IC) generation (Section~\ref{sec:ics}), zoom-in simulation settings and runs (Section~\ref{sec:zoom-in}), and post-processing and analysis (Section~\ref{sec:post-process}). Tables~\ref{tab:summary} summarizes our simulations and SHMF suppression results.

\subsection{Generating ICs}
\label{sec:ics}

We use the publicly available \textsc{CLASS} \citep{class}\footnote{\url{https://github.com/lesgourg/class_public/tree/master}} code to generate the matter density transfer functions $T_{\delta,\mathrm{CDM}}$ and velocity transfer functions $T_{\theta,\mathrm{CDM}}$ in CDM, as matter density and velocity ICs at redshift of $z=99$. Then, we compute the density and velocity transfer functions for f-NCDM models by multiplying the CDM transfer function by our transfer function parameterization from Equation~\ref{eq:t_fncdm_final}:  
\begin{equation}
    T_{\delta,\mathrm{f-NCDM}} = \mathcal{T}_{\mathrm{f-NCDM}}\times T_{\delta, \mathrm{CDM}}, \ \ \ \ 
    T_{\theta,\mathrm{f-NCDM}} = \mathcal{T}_{\mathrm{f-NCDM}}\times T_{\theta, \mathrm{CDM}}.\label{eq:tk_fncdm}
\end{equation}
We use Equation~\ref{eq:tk_fncdm} to generate transfer functions for the $24$ models summarized in Table~\ref{tab:summary}, which range from $k_{\mathrm{hm}}=22.8$ to $95.3~\mathrm{Mpc}^{-1}$ (or $m_{\mathrm{WDM}}=3$--$10~\mathrm{keV}$) for $\delta\in [0.2,~ 0.4,~ 0.6,~ 0.8]$. In addition, we use results from the $100\%$ WDM simulations in \citetalias{Nadler241003635}, which correspond to $\delta=0$. Figure~\ref{fig:transfers} shows transfer functions for the f-NCDM simulations presented in this work.

We use these transfer functions as the inputs for \textsc{MUSIC}\footnote{\url{https://www-n.oca.eu/ohahn/MUSIC/}} \citep{Hahn11036031} to generate the zoom-in ICs. We neglect thermal velocities when initializing our simulations. For the MW-like host Halo004 we zoom in on, we initialize a region at $z=99$ that corresponds to the Lagrangian volume of particles within $10$ times the virial radius of the host halo in the parent box at $z=0$; see \cite{Nadler220902675} for a detailed description of the parent box, and \cite{Buch240408043} for details on Halo004. Our simulations use four refinement regions relative to the parent box, yielding an equivalent of $8192$ particles per side in the highest-resolution region. For these fiducial-resolution simulations, the DM particle mass in the highest-resolution regions is $m_{\mathrm{part}}=4.0\times 10^5~M_{\mathrm{\odot}}$. In Appendix~\ref{sec:convergence}, we present convergence tests using four higher-resolution simulations for $k_{\mathrm{hm}}=22.8~\mathrm{Mpc}^{-1}$ and $\delta\in [0.2,~0.4,~0.6,~0.8]$.

\begin{figure}[t!]
\centering
\hspace{-5mm}
\includegraphics[width=0.46\textwidth]{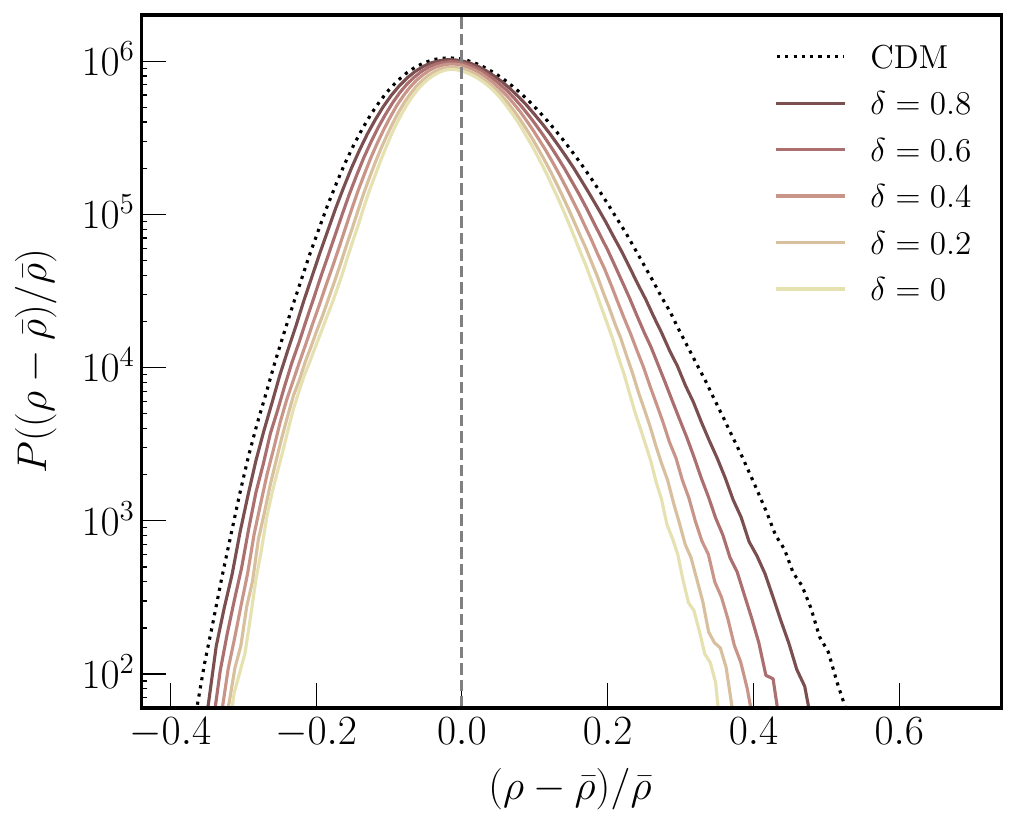}
    \caption{Distribution of local density contrast for HR particles in our zoom-in region at $z=99$ in CDM (dotted black line) and in f-NCDM models with $k_{\mathrm{hm}}=22.8~\mathrm{Mpc}^{-1}$ and transfer function plateau heights from $\delta=0.8$ to $0$ (dark to light solid lines). The dashed gray vertical line separates over and underdensities. Density contrasts are computed using \textsc{pynbody} \citep{pynbody}.}
    \label{fig:delta_ics}
\end{figure}

For each f-NCDM model, we pass the transfer functions described above into \textsc{MUSIC}, using its plug-in for \textsc{CAMB} transfer functions and the \textsc{CLASS}--\textsc{CAMB} conversion described in \citetalias{Nadler241003635}. We use the same random seeds when generating the refinement regions; thus, the phases of density modes are fixed in our f-NCDM simulations, and the only difference relative to CDM is that the amplitude of each mode is multiplied by $\mathcal{T}^2_{\mathrm{f-NCDM}}(k)$. Figure~\ref{fig:delta_ics} shows the distribution of local density contrast $(\rho-\bar{\rho})/\bar{\rho}$, where $\bar{\rho}$ is the mean matter density in the zoom-in region of our fiducial-resolution simulations at $z=99$, in CDM and in f-NCDM models with $k_{\mathrm{hm}}=22.8~\mathrm{Mpc}^{-1}$ (corresponding to $m_\mathrm{WDM}=3~\mathrm{keV}$). The f-NCDM overdensity distributions cut off at both large positive and negative density contrasts, indicating that small-scale overdensities and underdensities (corresponding to low-mass halos and small voids) are suppressed, relative to CDM. Larger f-NCDM components (i.e., smaller values of $\delta$) lead to more pronounced suppression. Figure~\ref{fig:delta_ics} is qualitatively representative of all f-NCDM models we consider.

\subsection{Zoom-in Simulations}
\label{sec:zoom-in}

\begin{deluxetable*}{{cccccc}}[t!]
\centering
\tablecolumns{6}
\tablecaption{Summary of COZMIC II Simulations}
\tablehead{\colhead{Scenario} & \colhead{$k_{\mathrm{hm}}~[\mathrm{Mpc}^{-1}]$} & \colhead{$M_{\mathrm{hm}}~[M_{\mathrm{\odot}}]$} & \colhead{$m_{\mathrm{WDM}}~[\mathrm{keV}]$} & \colhead{$\left.\left(\frac{N_{\mathrm{f-NCDM}}}{N_{\mathrm{CDM}}}\right)\right|_{1.2\times 10^8~M_{\mathrm{\odot}}}$} & \colhead{Color and Linestyle}}

\startdata 
\hline \hline
CDM &
-- &
-- &
1.0 &
\begin{tikzpicture}[yscale=0.5] \draw [line width=0.45mm,dotted,black] (0,-1) -- (1,-1) node[right]{};; \end{tikzpicture}
\\
%%%
\hline
%%%
\phantom{.} & 
22.8 & 
$4.3\times 10^8$ &
3 & 
0.85 &
\begin{tikzpicture}[yscale=0.5] \draw [line width=0.25mm,wdm_3_delta_08] (0,-1) -- (1,-1) node[right]{};; \end{tikzpicture}
\\
\phantom{.} & 
32.1 & 
$1.5\times 10^8$ &
4 & 
0.92 & 
\begin{tikzpicture}[yscale=0.5] \draw [line width=0.25mm,wdm_4_delta_08] (0,-1) -- (1,-1) node[right]{};; \end{tikzpicture}
\\
$\delta=0.8$ & 
41.8 &
$7.0\times 10^7$ &
5 & 
0.95 &
\begin{tikzpicture}[yscale=0.5] \draw [line width=0.25mm,wdm_5_delta_08] (0,-1) -- (1,-1) node[right]{};; \end{tikzpicture}
\\
($f_{\mathrm{NCDM}}=0.2$) & 
52.0 & 
$3.6\times 10^7$ &
6 & 
1.0 & 
\begin{tikzpicture}[yscale=0.5] \draw [line width=0.25mm,wdm_6_delta_08] (0,-1) -- (1,-1) node[right]{};; \end{tikzpicture}
\\
\phantom{.} & 
57.1 & 
$2.7 \times 10^7$ &
6.5 & 
1.0 & 
\begin{tikzpicture}[yscale=0.5] \draw [line width=0.25mm,wdm_6_5_delta_08] (0,-1) -- (1,-1) node[right]{};; \end{tikzpicture}
\\
\phantom{.} & 
95.3 & 
$5.9\times 10^6$ &
10 &
1.0 &
\begin{tikzpicture}[yscale=0.5] \draw [line width=0.25mm,wdm_10_delta_08] (0,-1) -- (1,-1) node[right]{};; \end{tikzpicture}
\\
%%%
\hline
%%%
\phantom{.} & 
22.8 & 
$4.3\times 10^8$ &
3 & 
0.75 &
\begin{tikzpicture}[yscale=0.5] \draw [line width=0.25mm,wdm_3_delta_06] (0,-1) -- (1,-1) node[right]{};; \end{tikzpicture}
\\
\phantom{.} & 
32.1 & 
$1.5\times 10^8$ &
4 & 
0.90 & 
\begin{tikzpicture}[yscale=0.5] \draw [line width=0.25mm,wdm_4_delta_06] (0,-1) -- (1,-1) node[right]{};; \end{tikzpicture}
\\
$\delta=0.6$ & 
41.8 &
$7.0\times 10^7$ &
5 & 
0.94 &
\begin{tikzpicture}[yscale=0.5] \draw [line width=0.25mm,wdm_5_delta_06] (0,-1) -- (1,-1) node[right]{};; \end{tikzpicture}
\\
($f_{\mathrm{NCDM}}$=0.4) & 
52.0 &
$3.6\times 10^7$ &
6 & 
0.91 & 
\begin{tikzpicture}[yscale=0.5] \draw [line width=0.25mm,wdm_6_delta_06] (0,-1) -- (1,-1) node[right]{};; \end{tikzpicture}
\\
\phantom{.} & 
57.1 & 
$2.7 \times 10^7$ &
6.5 & 
0.96 & 
\begin{tikzpicture}[yscale=0.5] \draw [line width=0.25mm,wdm_6_5_delta_06] (0,-1) -- (1,-1) node[right]{};; \end{tikzpicture}
\\
\phantom{.} & 
95.3 &
$5.9\times 10^6$ &
10 &
1.0 &
\begin{tikzpicture}[yscale=0.5] \draw [line width=0.25mm,wdm_10_delta_06] (0,-1) -- (1,-1) node[right]{};; \end{tikzpicture}
\\
%%%
\hline
%%%
\phantom{.} & 
22.8 & 
$4.3\times 10^8$ & 
3 & 
0.63 &
\begin{tikzpicture}[yscale=0.5] \draw [line width=0.25mm,wdm_3_delta_04] (0,-1) -- (1,-1) node[right]{};; \end{tikzpicture}
\\
\phantom{.} & 
32.1 & 
$1.5\times 10^8$ &
4 & 
0.80 & 
\begin{tikzpicture}[yscale=0.5] \draw [line width=0.25mm,wdm_4_delta_04] (0,-1) -- (1,-1) node[right]{};; \end{tikzpicture}
\\
$\delta=0.4$ & 
41.8 &
$7.0\times 10^7$ &
5 & 
0.88 &
\begin{tikzpicture}[yscale=0.5] \draw [line width=0.25mm,wdm_5_delta_04] (0,-1) -- (1,-1) node[right]{};; \end{tikzpicture}
\\
($f_{\mathrm{NCDM}}=0.6$) & 
52.0 &
$3.6\times 10^7$ &
6 & 
0.92 & 
\begin{tikzpicture}[yscale=0.5] \draw [line width=0.25mm,wdm_6_delta_04] (0,-1) -- (1,-1) node[right]{};; \end{tikzpicture}
\\
\phantom{.} & 
57.1 &
$2.7 \times 10^7$ &
6.5 & 
0.92 & 
\begin{tikzpicture}[yscale=0.5] \draw [line width=0.25mm,wdm_6_5_delta_04] (0,-1) -- (1,-1) node[right]{};; \end{tikzpicture}
\\
\phantom{.} & 
95.3 &
$5.9\times 10^6$ &
10 &
1.0 &
\begin{tikzpicture}[yscale=0.5] \draw [line width=0.25mm,wdm_10_delta_04] (0,-1) -- (1,-1) node[right]{};; \end{tikzpicture}
\\
%%%
\hline
%%%
\phantom{.} & 
22.8 & 
$4.3\times 10^8$ & 
3 & 
0.57 &
\begin{tikzpicture}[yscale=0.5] \draw [line width=0.25mm,wdm_3_delta_02] (0,-1) -- (1,-1) node[right]{};; \end{tikzpicture}
\\
\phantom{.} & 
32.1 & 
$1.5\times 10^8$ &
4 & 
0.69 & 
\begin{tikzpicture}[yscale=0.5] \draw [line width=0.25mm,wdm_4_delta_02] (0,-1) -- (1,-1) node[right]{};; \end{tikzpicture}
\\
$\delta=0.2$ & 
41.8 &
$7.0\times 10^7$ &
5 & 
0.81 &
\begin{tikzpicture}[yscale=0.5] \draw [line width=0.25mm,wdm_5_delta_02] (0,-1) -- (1,-1) node[right]{};; \end{tikzpicture}
\\
($f_{\mathrm{NCDM}}=0.8$) & 
52.0 &
$3.6\times 10^7$ &
6 & 
0.91 & 
\begin{tikzpicture}[yscale=0.5] \draw [line width=0.25mm,wdm_6_delta_02] (0,-1) -- (1,-1) node[right]{};; \end{tikzpicture}
\\
\phantom{.} & 
57.1 & 
$2.7 \times 10^7$ &
6.5 & 
0.95 & 
\begin{tikzpicture}[yscale=0.5] \draw [line width=0.25mm,wdm_6_5_delta_02] (0,-1) -- (1,-1) node[right]{};; \end{tikzpicture}
\\
\phantom{.} & 
95.3 &
$5.9\times 10^6$ &
10 &
0.99 &
\begin{tikzpicture}[yscale=0.5] \draw [line width=0.25mm,wdm_10_delta_02] (0,-1) -- (1,-1) node[right]{};; \end{tikzpicture}
\\
\hline \hline
\enddata
{\footnotesize \tablecomments{The first column lists the transfer function plateau amplitude and corresponding non-CDM fraction for each f-NCDM model we simulate, the second (third) column lists the input suppression scale $k_{\mathrm{hm}}$ ($M_{\mathrm{hm}}$) used to generate transfer functions and ICs for our simulations, the third column lists the corresponding thermal-relic WDM mass $m_{\mathrm{WDM}}$ that produces an onset of power suppression at the same wavenumber, the fourth column lists the suppression of the cumulative SHMF evaluated at our present-day virial mass resolution threshold of $M_{\mathrm{sub}}=1.2\times 10^8~M_{\mathrm{\odot}}$, and the fifth column shows the color and linestyle used for each model throughout this paper.}}
\label{tab:summary}
\end{deluxetable*}
%Using this procedure, we generate the transfer functions for $k_{\mathrm{hm}}\in [22.8,~ 32.1,~ 41.8,~ 52.0,~ 57.1,~ 95.3]~\mathrm{Mpc}^{-1}$, corresponding to $m_{\mathrm{WDM}}\in [3,~ 4,~ 5,~ 6,~ 6.5,~ 10]~\mathrm{keV}$, and $\delta\in [0.2,~ 0.4,~ 0.6,~ 0.8]$, for a total of $24$ fiducial-resolution simulations. Equivalently, we simulate $\alpha\in [18.2,~ 12.9,~ 9.9,~ 8.0,~ 7.3,~ 4.3]~\mathrm{kpc}$ with non-CDM fractions of $f_{\mathrm{NCDM}}\in [0.8,~ 0.6,~ 0.4,~ 0.2]$. In addition, we use results from the $100\%$ WDM simulations in \citetalias{Nadler241003635}, which correspond to $\delta=0$. Figure~\ref{fig:transfers} shows transfer functions for the f-NCDM simulations presented in this work.

\begin{figure*}[t!]
\centering
\hspace{-1cm}
\includegraphics[width=\textwidth]{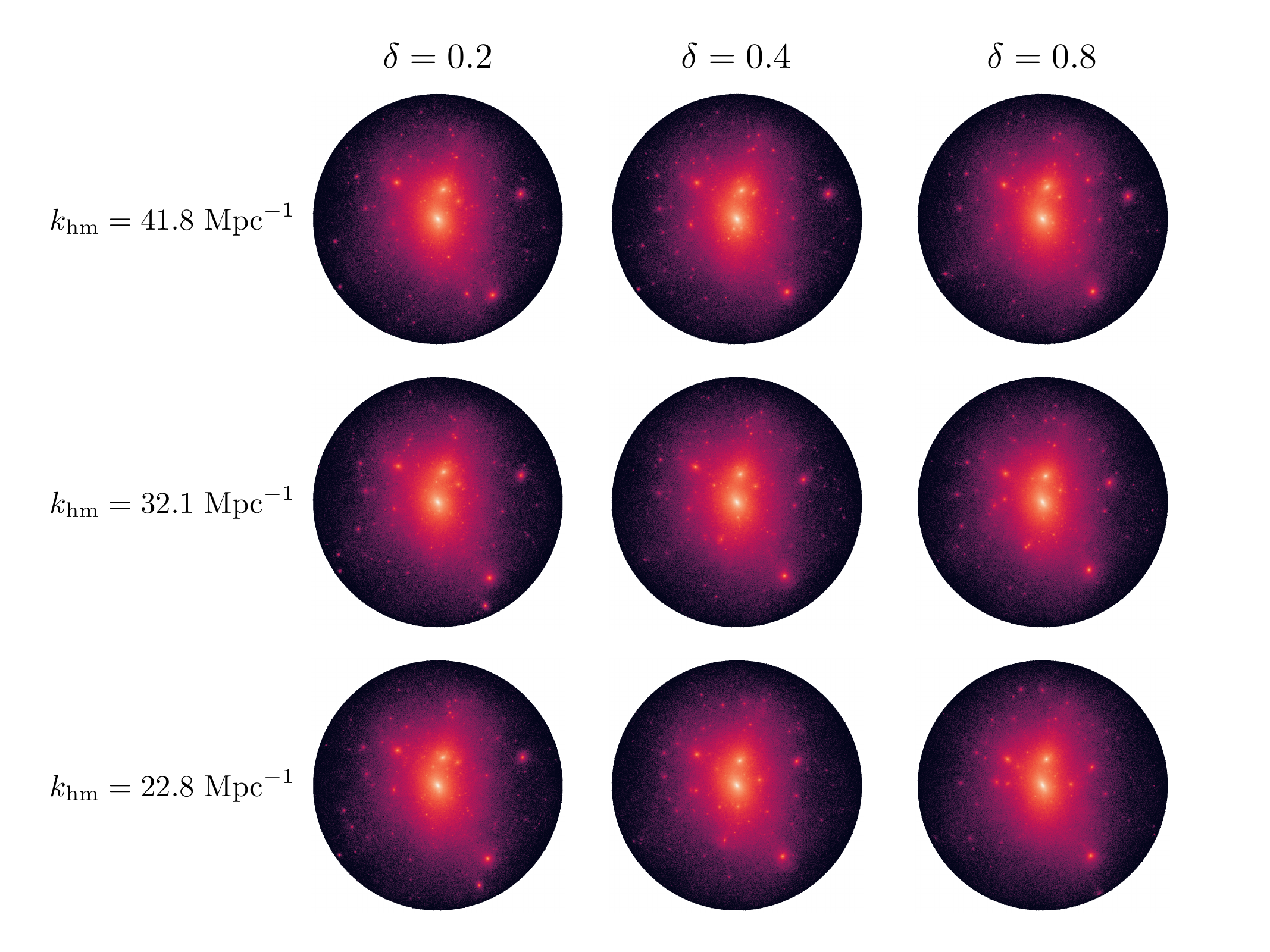}
    \caption{Projected DM density maps for several representative examples of our f-NCDM simulations of an MW-like system. The left column and top row, respectively, show the characteristic scales describing the suppression of the corresponding model's transfer function: the wavenumber $k_{\mathrm{hm}}$ and the plateau height $\delta$. Each visualization is centered on the host halo and spans $1.5$ times its virial radius. Visualizations were created using \textsc{meshoid} (\url{https://github.com/mikegrudic/meshoid}).}
    \label{fig:vis_main}
\end{figure*}

We run zoom-in simulations using \textsc{Gadget-2} \citep{Springel0505010} with a Plummer-equivalent gravitational softening of $\epsilon=170~\mathrm{pc}~h^{-1}$. Figure~\ref{fig:vis_main} shows projected DM density maps from our fiducial-resolution simulations for several f-NCDM suppression scales and plateau heights, focusing on a region spanning $1.5$ times the zoom-in host's virial radius. Small-scale structure is suppressed in the f-NCDM simulations in a manner that depends on the shape of the suppression (scale and plateau height). For example, for full NCDM scenarios with $\delta_{\mathrm{WDM}} = 0$ there is very little visible substructure for the $k_{\mathrm{hm}}=22.8~\mathrm{Mpc}^{-1}$ case (corresponding to $m_{\mathrm{WDM}}=3~\mathrm{keV}$), while the $k_{\mathrm{hm}}=41.8~\mathrm{Mpc}^{-1}$ case (corresponding to $m_{\mathrm{WDM}}=5~\mathrm{keV}$) is visually similar to CDM for large values of $\delta$. Meanwhile, for fixed $k_{\mathrm{hm}}$, the small-scale structure is visibly less suppressed for larger values of $\delta$, as expected.

\subsection{Post-processing and Analysis}
\label{sec:post-process}

We generate halo catalogs and merger trees by running {\sc Rockstar} and {\sc consistent-trees} \citep{Behroozi11104372,Behroozi11104370} on the HR particles from each simulation's $236$ output snapshots, which are evenly spaced in $\log(a)$, where $a=1/(1+z)$ is the scale factor, and redshifts range from $z\approx 10$ to $z=0$. We analyze all simulations at $z=0$, leaving an analysis of f-NCDM structure growth to future work. We measure SHMFs using peak virial mass $M_{\mathrm{sub,peak}}\equiv \max(M_{\mathrm{sub}}(z))$, because $M_{\mathrm{sub,peak}}$ more directly relates to the scale of the linear density perturbation that collapsed into a given halo compared to the present-day (stripped) subhalo mass.

When measuring SHMFs, we always apply a cut on present-day subhalo virial mass of $M_{\mathrm{sub}}(z=0)>1.2\times 10^8~M_{\mathrm{\odot}}$. In Appendix~\ref{sec:convergence}, we show that the f-NCDM SHMF suppression is converged down to this present-day mass threshold. Based on the results in \citetalias{Nadler241003635}, spurious subhalos formed through artificial fragmentation \citep{Wang0702575} contribute negligibly to our subhalo populations above this mass cut for the suppression scales we simulate, given the resolution of our simulations. Artificial fragmentation is even less important for our f-NCDM simulations compared to $100\%$ beyond-CDM simulations since our transfer functions do not completely cut off. Thus, we do not discard any subhalos with $M_{\mathrm{sub}}(z=0)>1.2\times 10^8~M_{\mathrm{\odot}}$.

\section{SHMFs}
\label{sec:basic}

\begin{figure*}[t!]
\centering
\includegraphics[width=\textwidth]{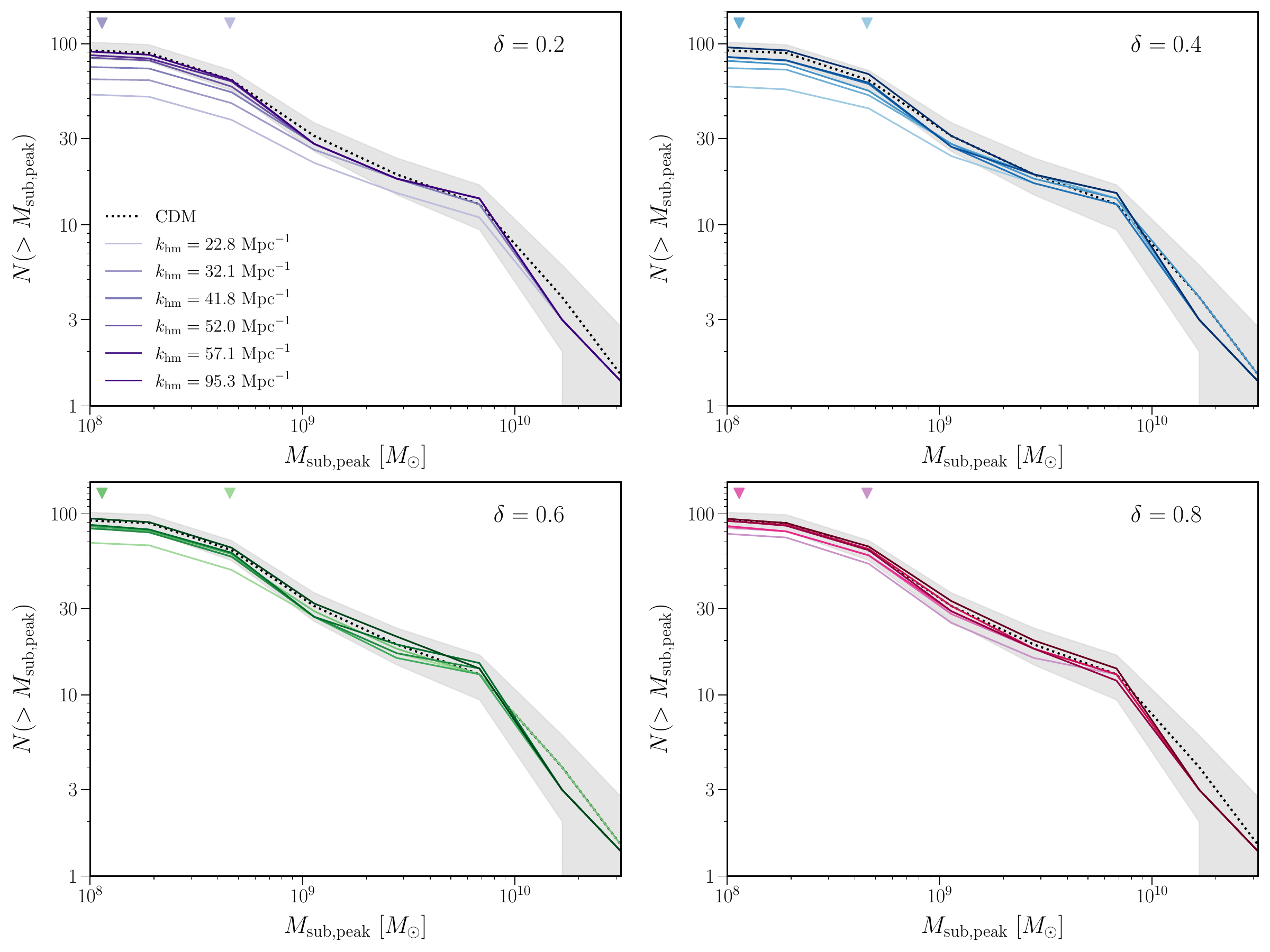}
    \caption{Cumulative SHMF, as a function of peak virial mass $M_{\mathrm{sub,peak}}$, for the MW-like host in  f-NCDM models (solid colored lines) and in CDM (dotted black lines). We simulate f-NCDM with $\delta = 0.2$ (top-left; purple), $0.4$ (top-right; blue), $0.6$ (bottom left panel; green), and $0.8$ (bottom right; pink); for each $\delta$, transfer function suppression scales range from $k_{\mathrm{hm}}=22.8$ (lightest) to $95.3~\mathrm{Mpc}^{-1}$ (darkest). All panels are restricted to subhalos with present-day virial masses $M_{\mathrm{sub}}>1.2\times 10^8~M_{\mathrm{\odot}}$, corresponding to $>300$ particles in the fiducial-resolution simulations. Markers indicate the half-mode masses of the $k_{\mathrm{hm}}=22.8$ and $32.1~\mathrm{Mpc}^{-1}$ models, in each panel; the half-mode masses for models with smaller suppression scales are outside the range of the plot. Gray bands show the $1\sigma$ Poisson uncertainty on the cumulative CDM SHMF.}
    \label{fig:shmf}
\end{figure*}

Figure~\ref{fig:shmf} shows SHMFs for the MW-like host in all f-NCDM models we simulate. These cumulative SHMFs flatten at low $M_{\mathrm{sub,peak}}$, even in CDM, because of our present-day $M_{\mathrm{sub}}$ cut. Note that the host-to-host variance among MW systems is comparable to the Poisson uncertainty we measure at $M_{\mathrm{sub,peak}}\approx 10^8~M_{\mathrm{\odot}}$ \citep{Mao150302637}.

SHMFs are clearly suppressed in our f-NCDM simulations. For $\delta=0.2$ and $0.4$, this suppression is significant relative to the Poisson uncertainty at $M_{\mathrm{sub,peak}}\approx 10^8~M_{\mathrm{\odot}}$, for most of the f-NCDM models we simulate. For $\delta=0.6$, the suppression is only significant for $k_{\mathrm{hm}}=22.8~\mathrm{Mpc}^{-1}$; for $\delta=0.8$, the suppression is not significant for any suppression scale. For the scenarios with the smallest $k_{\mathrm{hm}}$ and $\delta$ we simulate, subhalo abundances are significantly suppressed up to $M_{\mathrm{sub,peak}}\approx 10^{9}~M_{\mathrm{\odot}}$, above the half-mode mass of their corresponding WDM models (see the markers in Figure~\ref{fig:shmf}). Thus, the amount of SHMF suppression is set by both $k_{\mathrm{hm}}$ and $\delta$.

%%%%%%%%%%%%%%%%%%%%%%
%%%%%%%%%%%%%%%%%%%%%%

\section{SHMF Modeling}
\label{sec:shmf}

\subsection{Model}

To model the suppression of the SHMF in f-NCDM simulations, we fit an SHMF suppression model analogous to that in \citetalias{Nadler241003635}, using the likelihood framework described there. We briefly discuss the key components of the model and likelihood framework here.

We write the differential SHMF in the f-NCDM model $i$ as
\begin{align}
    \frac{\mathrm{d}N_{ij}}{\mathrm{d}\log(M_{\mathrm{sub,peak}})} &= \left(\frac{a}{100}\right)\left(\frac{M_{\mathrm{sub,peak}}}{M_{\mathrm{host,}ij}}\right)^{-b}e^{-50\left(\frac{M_{\mathrm{sub,peak}}}{M_{\mathrm{host,}i}}\right)^4}\nonumber& \\ &\times p_{i}(M_{\mathrm{sub}}>M_{\mathrm{min}}\lvert M_{\mathrm{sub,peak}})&\nonumber \\ &\times f_{\mathrm{f-NCDM}}(M_{\mathrm{sub,peak}},\theta_i,\alpha,\beta,\gamma),&\label{eq:shmf_model}
\end{align}
where $a$ ($b$) is the SHMF normalization (slope); $p_i(M_{\mathrm{sub}}>M_{\mathrm{min}}\lvert M_{\mathrm{sub,peak}})$ is the fraction of subhalos stripped below a present-day mass of $M_{\mathrm{min}}=1.2\times 10^8~M_{\mathrm{\odot}}$; $f_{\mathrm{f-NCDM}}(M_{\mathrm{sub,peak}},\theta_i,\alpha,\beta,\gamma)$ is the SHMF suppression; $\theta_i$ are the f-NCDM parameters; and $\alpha$, $\beta$, and $\gamma$ characterize the SHMF suppression amplitude and shape. We measure $p_i(M_{\mathrm{sub}}>M_{\mathrm{min}}\lvert M_{\mathrm{sub,peak}})$ by fitting a kernel density estimate (KDE) to the $M_{\mathrm{sub}}$--$M_{\mathrm{sub,peak}}$ relation in each f-NCDM simulation, and integrating each KDE above $M_{\mathrm{min}}$.

We fit the SHMF for each $\delta$ separately. Thus, for each $\delta$, $\theta_i$ spans the six $k_{\mathrm{hm}}$ values we simulate. We model $f_{\mathrm{f-NCDM}}$ as
\begin{align}
    f_{\mathrm{f-NCDM}}(M_{\mathrm{sub,peak}}&,M_{\mathrm{hm}},\alpha,\beta,\gamma)&\nonumber \\ &= (1-\delta)\times \left[1+\left(\frac{\alpha M_{\mathrm{hm}}}{M_{\mathrm{sub,peak}}}\right)^{\beta}\right]^{-\gamma} + \delta.&\label{eq:f_NCDM}
\end{align}
Here, $\alpha$ controls the mass scale and amplitude of the SHMF suppression, while $\beta$ and $\gamma$ control its shape above and below this scale. This form generalizes the SHMF suppression adopted for $100\%$ WDM \citep{Lovell13081399,Benito200111013} and assumes that SHMF suppression plateaus to the same amplitude as the transfer function, which we show accurately describes our simulation results. Note that Equation~\ref{eq:f_NCDM} reduces to the \citetalias{Nadler241003635} model for $100\%$ beyond-CDM scenarios for $\delta=0$, and that $f_{\mathrm{f-NCDM}}$ represents the f-NCDM SHMF suppression relative to CDM.

For each $\delta$, we fit the parameters of our SHMF model---$\vec{\Theta}=\{a,~b,~\alpha,~\beta,~\gamma\}$---to the f-NCDM simulation results, using the Poisson likelihood from \citetalias{Nadler241003635}. We use linear uniform priors for $0.1<a<10$ and $0<\alpha<50$, and Jeffreys priors for $0.1<b<1.5$, $0<\beta<10$, and $0<\gamma<5$. We sample each five-dimensional posterior by running \textsc{emcee} \citep{emcee} for $10^5$ steps with $100$ walkers, discarding $10^4$ burn-in steps, which yields well-converged posteriors with hundreds of independent samples.

\subsection{f-NCDM Results}

We now summarize the results of our SHMF fits. We focus on $\delta=0.2$ and $0.4$ because these cases are constrained by the MW satellite analysis in Section~\ref{sec:limits}; we present results for $\delta=0.6$ and $0.8$ in Appendix~\ref{sec:other_delta}. Throughout, we show marginalized posteriors for $\alpha$, $\beta$, and $\gamma$; posteriors for $a$ and $b$ are similar to those in \citetalias{Nadler241003635} for Halo004, and are consistent between our fits across different values of $\delta$.

First, for $\delta=0.2$, we obtain $\alpha=3.5^{+17.0}_{-3.3}$, $\beta=0.9^{+0.8}_{-0.4}$, and $\gamma=0.2^{+0.4}_{-0.2}$, at $68\%$ confidence. Similar to the $100\%$ WDM and FDM results in \citetalias{Nadler241003635}, only $\beta$ is well measured, although we place upper limits on $\alpha$ and $\gamma$. The parameters that maximize the posterior for $\delta=0.2$ are
\begin{align}
    \alpha_{\delta=0.2}=5.8,\nonumber\\
    \beta_{\delta=0.2}=1.0,\nonumber\\
 \gamma_{\delta=0.2}=0.6.\label{eq:best_fit_delta_02}
\end{align}

Next, for $\delta=0.4$, we obtain $\alpha=2.3^{+11.5}_{-2.2}$, $\beta=1.1^{+1.1}_{-0.3}$, and $\gamma=0.3^{+0.6}_{-0.3}$, at $68\%$ confidence. Thus, the uncertainties on the SHMF suppression shape parameters are larger than in the $\delta=0.2$ case, while uncertainties for $\alpha$ are similar. The parameters that maximize the posterior for $\delta=0.4$ are
\begin{align}
    \alpha_{\delta=0.4}=4.7,\nonumber\\
    \beta_{\delta=0.4}=1.3,\nonumber\\
 \gamma_{\delta=0.4}=0.7.\label{eq:best_fit_delta_04}
\end{align}

\begin{figure}[t!]
\centering
\hspace{-5mm}
\includegraphics[trim={0 0.35cm 0 0cm},width=0.49\textwidth]{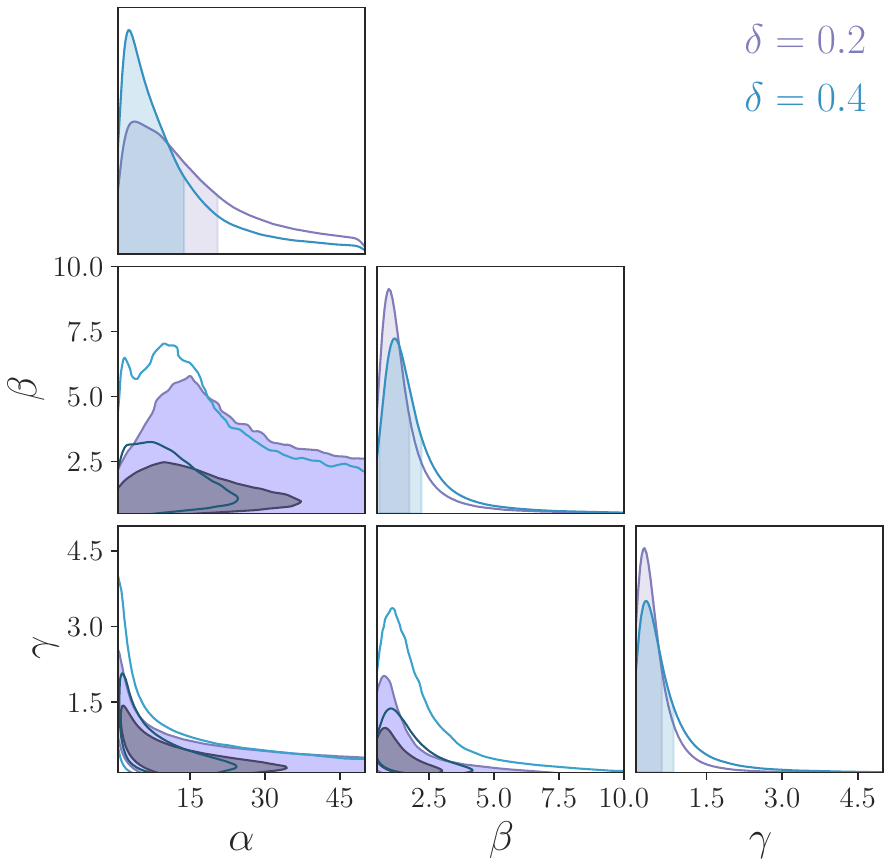}
    \caption{Marginalized posterior probability distribution for the f-NCDM SHMF suppression fit for $\delta=0.2$ (filled purple) and $0.4$ (unfilled blue). Dark (light) two-dimensional contours show $68\%$ ($95\%$) confidence intervals. Top and side panels show 1-d marginal posteriors with shaded $68\%$ confidence intervals.}
    \label{fig:shmf_posterior}
\end{figure}

Figure~\ref{fig:shmf_posterior} compares the marginalized posteriors for $\delta=0.2$ and $0.4$. These posteriors are consistent with each other and with the $100\%$ WDM posteriors from \citetalias{Nadler241003635}, which lends confidence to our f-NCDM SHMF suppression parameterization. In particular, for each $\delta$, our simple modification to the $100\%$ beyond-CDM SHMF suppression in Equation~\ref{eq:f_NCDM} accurately captures the SHMF in f-NCDM models---with goodness-of-fit statistics quantified below---without requiring significantly different values for $\alpha$, $\beta$, and $\gamma$.

\begin{figure*}[t!]
\includegraphics[width=0.5\textwidth]{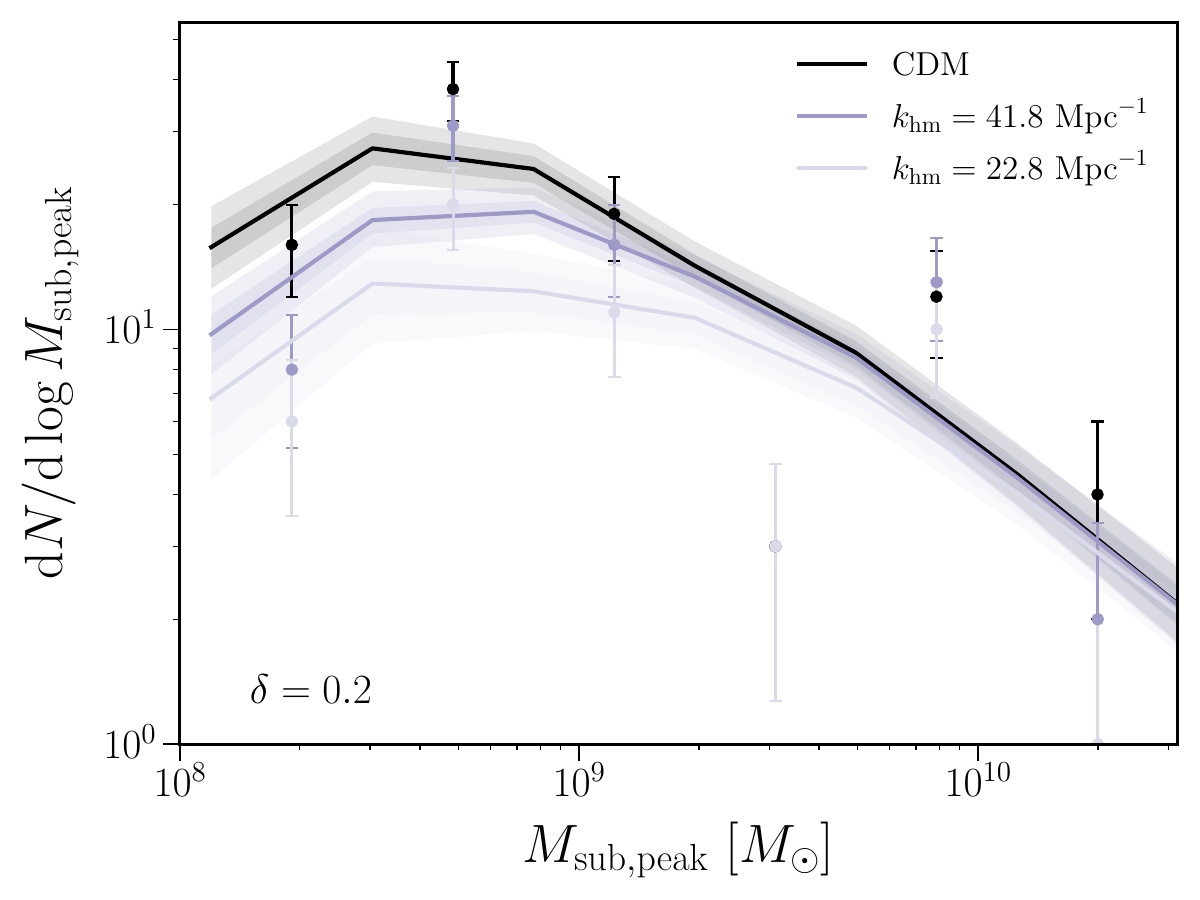}
\includegraphics[width=0.5\textwidth]{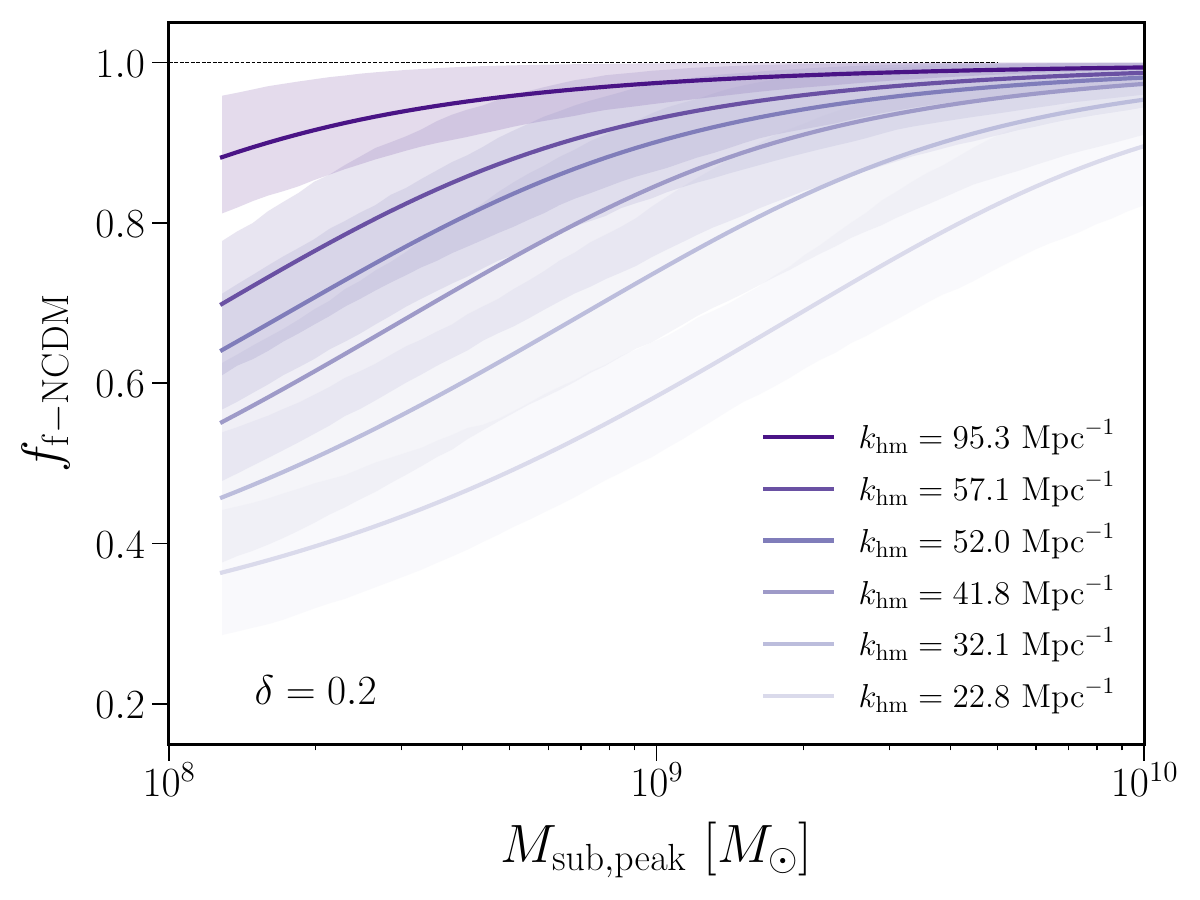}\\
\includegraphics[width=0.5\textwidth]{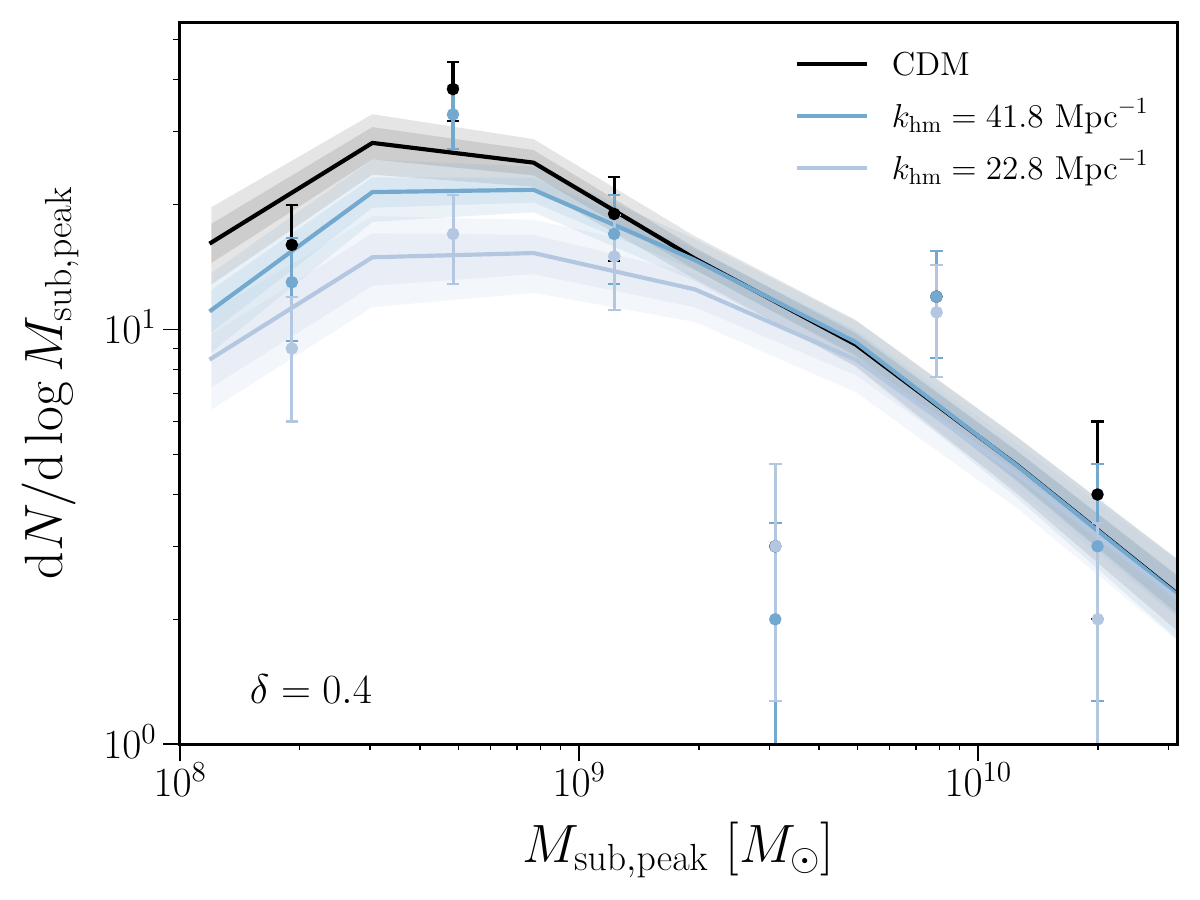}
\includegraphics[width=0.5\textwidth]{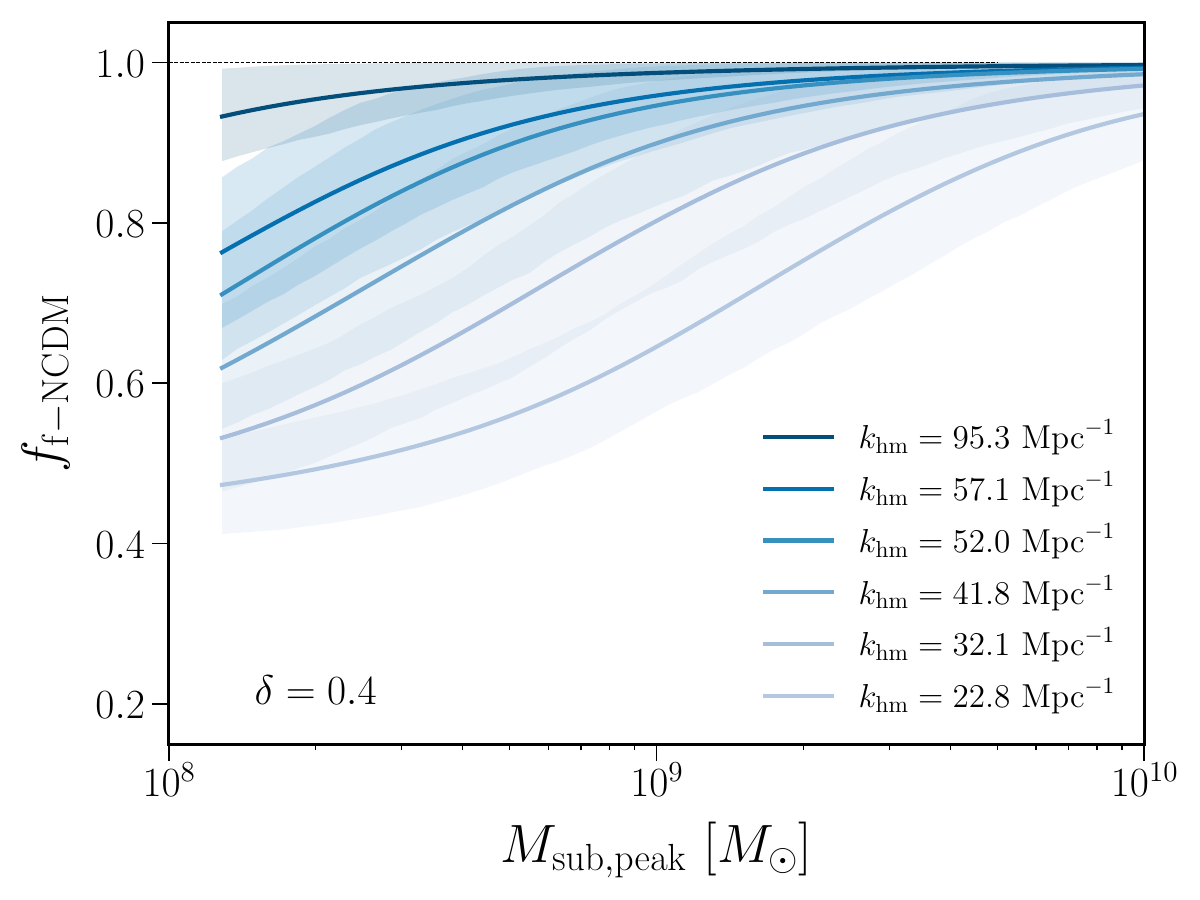}
    \caption{Left: Mean differential SHMF, as a function of peak virial mass, for f-NCDM simulations with $\delta=0.2$ (top; purple), $\delta=0.4$ (bottom; blue), and CDM (black). Error bars show $1\sigma$ Poisson errors on the mean. Right: Corresponding SHMF suppression predictions. Dark (light) bands show $68\%$ ($95\%$) confidence intervals from the SHMF fit in all panels, as calculated by sampling from the posterior.}
    \label{fig:shmf_pred}
\end{figure*}

Figure~\ref{fig:shmf_pred} compares the differential SHMF (left) and SHMF suppression (right), for $\delta=0.2$ (top) and $0.4$ (bottom), as predicted by our model and measured in our simulations. There is a significant statistical uncertainty associated with our simulation measurements of the SHMF because we only simulate one MW-like host in this study. Nonetheless, the f-NCDM SHMF is significantly suppressed relative to CDM in the lowest-$M_{\mathrm{sub,peak}}$ bins, and our fits capture this suppression well. In particular, for $\delta=0.2$ ($0.4$), we calculate $p=0.33$ ($0.23$) from a two-sample Kolmogorov–Smirnov (KS) test, indicating no significant difference between the predicted SHMF and the simulation results.\footnote{We also calculate a reduced-$\chi^2$ statistic of $\chi^2=0.5$ ($0.6$) across all $k_{\mathrm{hm}}$ we simulate for $\delta=0.2$ ($0.4$), which indicates a good fit in both cases. However, these values are sensitive to the assumed Poisson error model because we only simulate one MW-like host, which is subject to large statistical fluctuations. Thus, we focus on the KS test as a goodness-of-fit metric.} 

The predicted SHMF suppression in the right column of Figure~\ref{fig:shmf_pred} is a smooth function of $M_{\mathrm{sub,peak}}$ for both $\delta=0.2$ and $0.4$. For $\delta=0.2$, the f-NCDM SHMF suppression shape is similar to, but slightly weaker than, predictions for $100\%$ WDM (e.g., from \citealt{Lovell13081399}). Meanwhile, for $\delta=0.4$, the f-NCDM SHMF suppression significantly weakens. Furthermore, for the smallest $k_{\mathrm{hm}}$ we simulate with $\delta=0.4$, the suppression noticeably flattens at low $M_{\mathrm{sub,peak}}$ due to the constant term in our f-NCDM SHMF model (Equation~\ref{eq:f_NCDM}). In Appendix~\ref{sec:other_delta}, we show that the f-NCDM SHMF is even less suppressed for $\delta=0.6$ and $0.8$, as expected.

\section{Limits from MW Satellites}
\label{sec:limits}

We now use the f-NCDM SHMF suppression fits to place new bounds on f-NCDM transfer functions and the thermal-relic f-WDM fraction and mass, by forward modeling the resulting MW satellite population.

\subsection{Inference Framework}

We use the MW satellite inference framework from \cite{Nadler200800022}, which is based on two DM-only zoom-in simulations of MW-like systems \citep{Mao150302637}, an empirical galaxy--halo connection model \citep{Nadler171204467,Nadler180905542,Nadler191203303}, and DES and PS1 MW satellite population observations and selection functions \citep{Drlica-Wagner191203302}. For each f-NCDM scenario (i.e., for each $\delta$), we simultaneously fit for all eight galaxy--halo connection parameters and for the half-mode mass using the best-fit values for the corresponding SHMF suppression model from Section~\ref{sec:shmf}. Propagating f-NCDM SHMF suppression uncertainties negligibly affects the resulting limits, since galaxy--halo connection uncertainties dominate the error budget. This procedure assumes that f-NCDM physics only affects subhalo abundances relative to CDM, without altering the form of the galaxy--halo connection (consistent with the results of \citealt{Despali250112439}) or the subhalo radial distribution. Note, however, that we marginalize over the full galaxy--halo connection model when deriving f-NCDM constraints.

We implement the same galaxy--halo connection parameter priors as \cite{Nadler200800022}, and we use a linear prior on $\log(M_{\mathrm{hm}}/M_{\mathrm{\odot}})$ over the interval $[7,10]$ for each $\delta$. We run \textsc{emcee} for $10^6$ steps with $36$ walkers, discarding $10^5$ burn-in steps; this yields well-converged posteriors with hundreds of independent samples in all f-NCDM scenarios.

\subsection{f-NCDM Limits}
\label{sec:fncdm_limit}

\begin{figure}[t!]
\hspace{-5mm}
\includegraphics[width=0.5\textwidth]{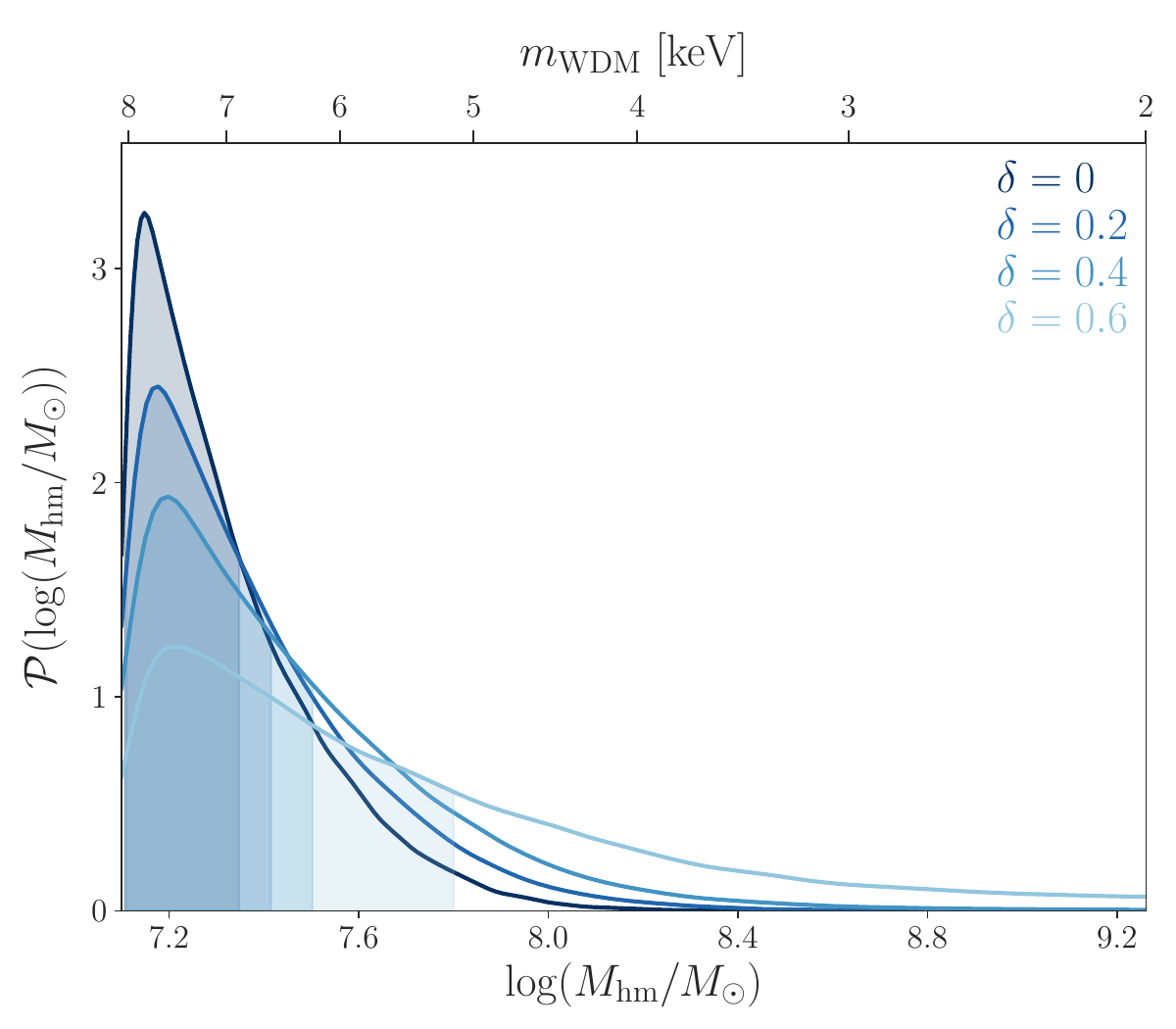}
    \caption{1-d marginalized posterior probability distributions for $\log(M_{\mathrm{hm}}/M_{\mathrm{\odot}})$, derived from our analysis of MW satellites, based on our new model of the f-NCDM SHMF. The results are shown for different transfer function plateau heights, $\delta=0.2$, $0.4$, and $0.6$ (dark to light blue). We also show the marginalized posterior using the $100\%$ WDM SHMF suppression fit from \citetalias{Nadler241003635} (darkest blue). Dark (light) regions beneath each posterior show $68\%$ ($95\%$) upper limits on $\log(M_{\mathrm{hm}}/M_{\mathrm{\odot}})$; note that we do not obtain a $95\%$ confidence upper limit for $\delta=0.6$.}
    \label{fig:all_frac_posterior}
\end{figure}

Figure~\ref{fig:all_frac_posterior} shows the marginalized $\log(M_{\mathrm{hm}}/M_{\mathrm{\odot}})$ posteriors for our $\delta=0.2$, $0.4$, and $0.6$ fits, from light to dark blue; we also show the corresponding posterior from the $100\%$ WDM fit in \citetalias{Nadler241003635} ($\delta=0$; darkest blue). The remaining galaxy--halo connection parameter constraints and degeneracies are similar to those in \cite{Nadler200800022} for $100\%$ WDM.

Thus, we derive the following $95\%$ confidence limits:
\begin{align}
    M_{\mathrm{hm}} &< 5.4\times 10^7~M_{\mathrm{\odot}},\ \delta=0.2,& \nonumber \\
    M_{\mathrm{hm}} &< 7.8\times 10^7~M_{\mathrm{\odot}},\ \delta = 0.4.&
\end{align}
These translate to lower limits of $k_{\mathrm{hm}}>46$ and $40~\mathrm{Mpc}^{-1}$, respectively. For reference, the $100\%$ WDM ($\delta=0$) constraint from \citetalias{Nadler241003635} is $M_{\mathrm{hm}}<3.9\times 10^7~M_{\mathrm{\odot}}$, corresponding to $k_{\mathrm{hm}}>51~\mathrm{Mpc}^{-1}$ at $95\%$ confidence. Thus, as expected, the constraints weaken as $\delta$ increases. Meanwhile, we do not obtain a $95\%$ confidence upper limit for $\delta=0.6$, because the weak SHMF suppression in this scenario does not yield a statistically significant underabundance of faint MW satellites, and all suppression scales are consistent with the data. Because the SHMF is strictly less suppressed for even larger values of $\delta$, we obtain an even flatter posterior for $\delta=0.8$, and omit it from Figure~\ref{fig:all_frac_posterior}, for clarity.

\begin{figure*}[t!]
\centering
\hspace{-5mm}
\includegraphics[width=\textwidth]{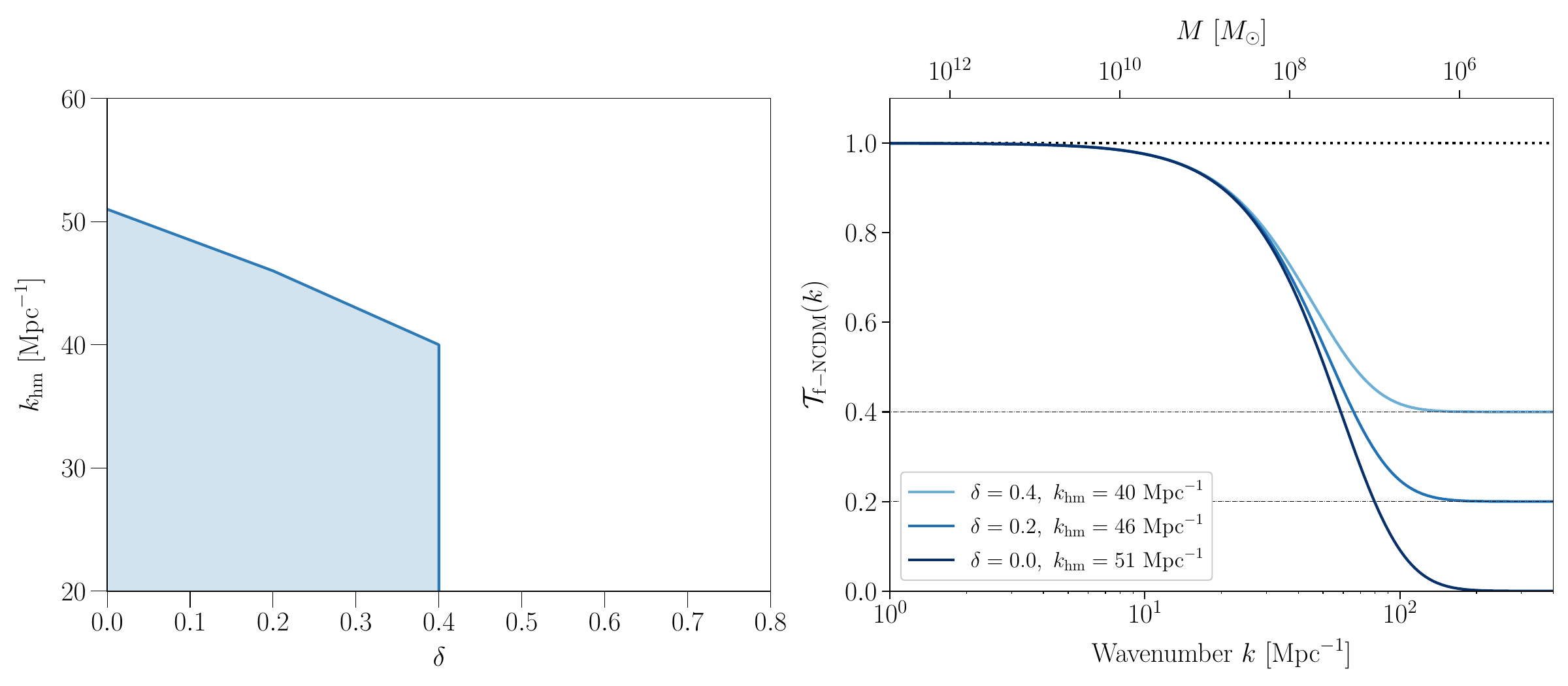}
    \caption{\emph{Left}: $95\%$ confidence lower limits on $k_{\mathrm{hm}}$, as a function of $\delta$, from our MW satellite analysis. The data are not sensitive to models with $\delta>0.4$, at $95\%$ confidence. \emph{Right}: The transfer functions corresponding to the bound shown in the left panel, evaluated at $\delta$ and $k_{\mathrm{hm}}$ values along the constraint. Note that the ruled-out transfer functions overlap for $k\lesssim 40~\mathrm{Mpc}^{-1}$.}
    \label{fig:constraints_on_khm}
\end{figure*}

The left panel of Figure~\ref{fig:constraints_on_khm} shows these constraints in the $k_{\mathrm{hm}}$--$\delta$ parameter space (where the shaded region is excluded at $95\%$ confidence). As discussed above, the lower limit on $k_{\mathrm{hm}}$ becomes more stringent with decreasing $\delta$, reaching the $100\%$ WDM result at $\delta=0$. The constraint does not exist above $\delta=0.4$, indicating that the DES and PS1 MW satellite luminosity population is not sensitive to models with weaker suppression in the transfer function than the $\delta=0.4$ model.

The right panel of Figure~\ref{fig:constraints_on_khm} shows f-NCDM transfer functions evaluated at the $95\%$ confidence-level bounds shown in the left panel; the $\delta=0$ result from \citetalias{Nadler241003635} is included for reference. The ruled-out transfer functions are nearly identical for wavenumbers smaller than $k\approx 40~\mathrm{Mpc}^{-1}$, which corresponds to the minimum mass of observed MW satellites \citep{Nadler191203303}. This is expected, since \cite{Nadler200800022} demonstrated that constraints on $100\%$ beyond-CDM models are driven by the abundance of the faintest observable satellites. Because the SHMF rises more steeply compared to the luminosity function, these galaxies occupy halos in a narrow range of peak virial mass, near $\approx 10^8~M_{\mathrm{\odot}}$, and are sourced by density fluctuations in a narrow wavenumber range near $k\approx 40~\mathrm{Mpc}^{-1}$. 

The shape of the transfer function suppression at $k\lesssim 40~\mathrm{Mpc}^{-1}$ can affect the abundance of subhalos with masses above $\approx 10^8~M_{\mathrm{\odot}}$ (e.g., \citealt{Stucker210909760}); in this regime, our ruled-out f-NCDM models have identical transfer functions. On the other hand, according to our f-NCDM SHMF parameterization in Equation~\ref{eq:f_NCDM}, higher transfer function plateaus at $k\gtrsim 40~\mathrm{Mpc}^{-1}$ lead to systematically weaker SHMF suppression at low masses. Thus, models with sufficiently large $\delta$ do not feature enough suppression in the SHMF (and thus also in the satellite luminosity function) to be constrained. Future observations of more faint satellite galaxies will reduce the statistical uncertainties on the faint-end luminosity function and better probe these f-NCDM scenarios (e.g., see the forecasts in \citealt{Nadler240110318}).

\subsection{f-WDM Limits}
\label{sec:fwdm_limit}

We now apply our f-NCDM transfer function limits to constrain f-WDM scenarios. In particular, we consider models where a fraction $f_{\mathrm{WDM}}$ of the total DM relic density is a thermal relic with mass $m_{\mathrm{WDM}}$, with the rest in CDM. We generate transfer functions for these models using \textsc{CLASS}. As discussed in Section~\ref{sec:model}, these f-WDM transfer functions do not plateau at large wavenumbers, unlike the f-NCDM transfer function parameterization we use to initialize the simulations. Nonetheless, we constrain f-WDM models by exploiting the similarity of their initial suppression to the ruled-out f-NCDM transfer functions presented above.

Specifically, we posit that f-WDM models featuring transfer functions that are strictly more suppressed than the ruled-out f-NCDM models from Section~\ref{sec:fncdm_limit} are inconsistent with the data; we then find the value of $k_{\mathrm{hm}}$ that gives the closest match to the slope of each f-WDM transfer function, for each $\delta$, and leverage the bound on the corresponding f-NCDM model to constrain the f-WDM scenarios. The results of this method are shown in the right panel of Figure~\ref{fig:constraints_on_khm}. Following \cite{Maamari201002936}, we require that the criterion above holds up to $k=91~\mathrm{Mpc}^{-1}$; this wavenumber is a factor of $\approx 2.2$ larger than the minimum halo mass wavenumber of $\approx 40~\mathrm{Mpc}^{-1}$, and therefore corresponds to an order-of-magnitude smaller halo mass. Thus, excess power relative to our ruled-out f-NCDM transfer functions at even larger wavenumbers than we consider in the constraint method, $k>91~\mathrm{Mpc}^{-1}$, is expected to have a negligible impact on the abundance of observable MW satellites. Note that the limits we derive using this approach are conservative because f-WDM transfer functions decrease more quickly on small scales than our ruled-out f-NCDM models.

\begin{figure*}[t!]
\centering
\includegraphics[width=\textwidth]{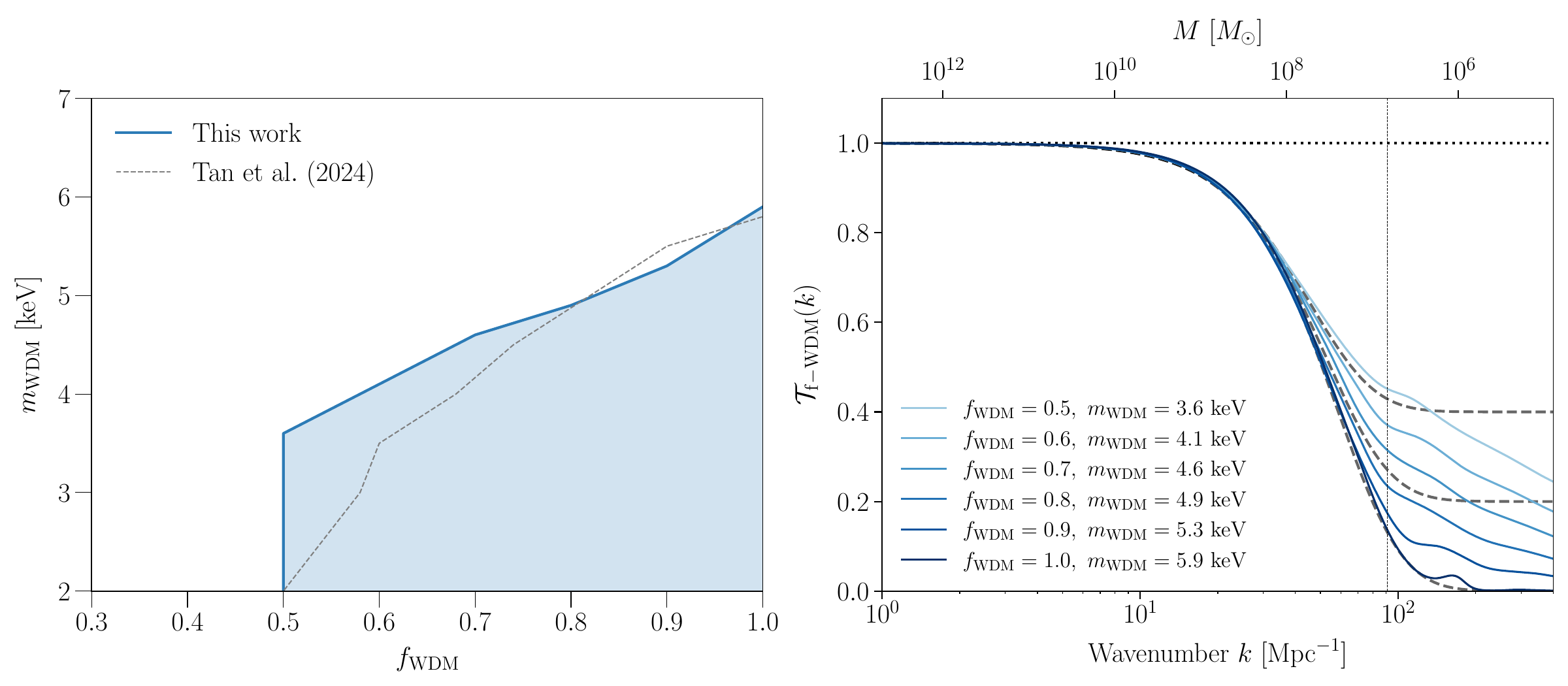}
    \caption{\emph{Left}: Lower bounds on WDM particle mass $m_{\mathrm{WDM}}$, as a function of WDM fraction $f_{\mathrm{WDM}}$; the shaded region is excluded at $95\%$ confidence. The bounds are inferred from analysis of MW satellite population, leveraging the forward model of the subhalo population in fractional non-CDM cosmologies introduced in this work. The dashed gray line shows the 20:1 odds ratio constraint from \cite{Tan240918917}, for comparison. We find that the current data are not sensitive to models with $f_{\mathrm{WDM}}<0.5$, at $95\%$ confidence. \emph{Right}: Transfer functions for WDM scenarios corresponding to the bounds on the left panel are shown by solid lines. The non-CDM transfer functions used to derive the bounds are shown by dashed lines and are evaluated along our constraint from Figure~\ref{fig:constraints_on_khm}. Fractional WDM constraints are derived by requiring that $\mathcal{T}_{\mathrm{f-WDM}}(k)$ is strictly more suppressed than the ruled-out dashed line for $k<91~\mathrm{Mpc}^{-1}$ (this boundary is marked by the vertical dashed line). We also show the $100\%$ WDM constraint from \citetalias{Nadler241003635}, $m_{\mathrm{WDM}}>5.9~\mathrm{keV}$ at $95\%$ confidence, for comparison.}
    \label{fig:constraints_on_wdm}
\end{figure*}

In this way, we derive the following $95\%$ confidence limits:
\begin{align}
    m_{\mathrm{WDM}} &> 3.6~\mathrm{keV},\ f_{\mathrm{WDM}}=0.5,& \nonumber \\
    m_{\mathrm{WDM}} &> 4.1~\mathrm{keV},\ f_{\mathrm{WDM}} = 0.6.& \nonumber \\
    m_{\mathrm{WDM}} &> 4.6~\mathrm{keV},\ f_{\mathrm{WDM}} = 0.7.& \nonumber \\
    m_{\mathrm{WDM}} &> 4.9~\mathrm{keV},\ f_{\mathrm{WDM}} = 0.8.& \nonumber \\
    m_{\mathrm{WDM}} &> 5.4~\mathrm{keV},\ f_{\mathrm{WDM}} = 0.9.& \label{eq:fwdm_limits}
\end{align}
These limits are shown in the left panel of Figure~\ref{fig:constraints_on_wdm}, where we also include the $100\%$ WDM case (i.e., $f_{\mathrm{WDM}}=1$) limit of $m_{\mathrm{WDM}}>5.9~\mathrm{keV}$ from \citetalias{Nadler241003635}. Similar to the limits on $k_{\mathrm{hm}}$, these bounds weaken for lower WDM fractions $f_{\mathrm{WDM}}$ and entirely disappear at $f_{\mathrm{WDM}}=0.5$. This limiting value of $f_{\mathrm{WDM}}$ is slightly smaller than the largest value of $\delta=0.4$ that we constrain due to the steeper transfer function suppression in f-WDM compared to our f-NCDM parameterization.

The right panel of Figure~\ref{fig:constraints_on_wdm} shows the f-WDM transfer functions that correspond to the upper bound on particle mass, for each $f_{\mathrm{WDM}}$ we consider. These f-WDM transfer functions match the corresponding ruled-out f-NCDM transfer functions at the percent level for $k<91~\mathrm{Mpc}^{-1}$, by construction, and continue to decrease on smaller scales rather than reaching a plateau. As a result, the corresponding SHMFs and satellite populations are expected to be indistinguishable, down to a minimum halo mass of $10^8~M_{\mathrm{\odot}}$, given the sensitivity of current MW satellite observations. Minor differences between the ruled-out f-WDM and f-NCDM transfer functions are visible due to the finite precision of our \textsc{CLASS} calculations; these differences negligibly affect our limits and do not change their interpretation. 

Our f-WDM analysis shows that constraints on a wide range of f-NCDM scenarios are enabled by our f-NCDM parameterization. In some scenarios---for example, models where a fractional component of DM interacts with Standard Model particles---will likely map even more precisely to our f-NCDM transfer functions. In other cases---including fractional ultralight DM models---both the transfer function cutoff shape and plateau may differ from our parameterization, and a more careful mapping and/or additional simulations will be needed to derive robust constraints. We leave these considerations to future work.

\section{Discussion}
\label{sec:discussion}

\subsection{Comparison to Previous Studies}

We are not aware of previous zoom-in simulations for the general f-NCDM transfer function parameterization we adopt. In particular, while \cite{Hooper2022} performed a grid of large-scale cosmological simulations to train a Ly$\alpha$ forest emulator using the same transfer function parameterization, it is challenging to directly compare our results to this work because of the different scales and environments covered by the respective simulations. We therefore focus on comparisons to f-WDM simulations and constraints.

In terms of simulations, \cite{Harada14121592} simulate f-WDM models with $f_{\mathrm{WDM}}\in[0.25,~ 0.5]$ and $m_{\mathrm{WDM}}=2.4~\mathrm{keV}$ in a $(5~\mathrm{Mpc}/h)^3$ cosmological volume with a slightly lower simulation particle mass than ours. For isolated halos, their mass functions are suppressed relative to CDM by $\approx 50\%$ and $25\%$ for $f_{\mathrm{WDM}}=0.5$ and $0.25$, respectively. We compare these to our $\delta=0.4$ and $0.8$ simulations with $k_{\mathrm{hm}}=22.8~\mathrm{Mpc}^{-1}$ (corresponding to $m_{\mathrm{WDM}}=3~\mathrm{keV}$), for which the SHMF suppression is $\approx 50\%$ and $20\%$ relative to CDM at the lowest masses we resolve (see Figure~\ref{fig:shmf_pred} and Appendix~\ref{sec:other_delta}). Thus, our results are consistent with those in \cite{Harada14121592}. We note that our SHMFs are slightly less suppressed than theirs, likely due to a combination of the higher $m_{\mathrm{WDM}}$ and less steeply decreasing transfer functions we simulate. In addition, although we measure SHMF rather than isolated halo mass function suppression, we expect our use of $M_{\mathrm{sub,peak}}$ to reduce differences between the two.

Meanwhile, \cite{Anderhalden12063788} and \cite{Parimbelli210604588}, respectively, performed zoom-in and cosmological f-WDM with significantly lower $m_{\mathrm{WDM}}$ than we consider. These studies demonstrate that, even for small $f_{\mathrm{WDM}}$, the SHMF and isolated halo mass function can be significantly suppressed relative to CDM. This is likely due to the decreasing small-scale power in f-WDM models, and suggests that even higher-resolution zoom-in simulations of the f-WDM models we constrain may reveal additional SHMF suppression at subhalo masses below $10^8~M_{\mathrm{\odot}}$. This regime will be probed by upcoming strong gravitational lensing and stellar stream observations (e.g., \citealt{Drlica-Wagner190201055})

In terms of constraints, a number of previous studies show that lower limits on $m_{\mathrm{WDM}}$ from small-scale structure observables significantly weaken as $f_{\mathrm{WDM}}$ decreases. As a result, these probes lose sensitivity to $m_{\mathrm{WDM}}$ below a minimum WDM fraction. For example, \cite{Boyarsky08120010} used Lyman-$\alpha$ forest measurements to show that $f_{\mathrm{WDM}}<0.4$ for $m_{\mathrm{WDM}}=1.1~\mathrm{keV}$, at $95\%$ confidence. \cite{Anderhalden12063788} used MW satellite abundances to
rule out extreme f-WDM models with $f_{\mathrm{WDM}}=0.2$ and $m_{\mathrm{WDM}}=0.1~\mathrm{keV}$, or $f_{\mathrm{WDM}}=0.5$ and $m_{\mathrm{WDM}}=0.3~\mathrm{keV}$. Meanwhile, \cite{Kamada160401489} concluded that gravitational lensing flux ratio anomaly measurements are consistent with any $m_{\mathrm{WDM}}$ for $f_{\mathrm{WDM}}<0.47$. Finally, \cite{Diamanti170103128} used MW satellite measurements from the Sloan Digital Sky Survey to show that $f_{\mathrm{WDM}}<0.29$ for fermions with $m_{\mathrm{WDM}}\in [1,~ 10]~\mathrm{keV}$, and $f_{\mathrm{WDM}}<0.43$ for fermions with $m_{\mathrm{WDM}}\in [10,~ 100]~\mathrm{keV}$.

Recently, \cite{Tan240918917} derived f-WDM limits by combining the semi-analytic code \textsc{SASHIMI} \citep{Dekker211113137}, f-WDM transfer functions, and the MW satellite inference framework used here and described in \cite{Nadler180905542,Nadler191203303,Nadler200800022}. The 20:1 odds ratio limit reported in that study on $f_{\mathrm{WDM}}$, as a function of $m_{\mathrm{WDM}}$, provides the closest comparison to our $95\%$ confidence limits, and is shown by the dashed gray line in the left panel of Figure~\ref{fig:constraints_on_wdm}. The result is in broad agreement with our constraints.

Thus, with the exception of the joint analysis in \cite{Diamanti170103128} and the recent results of \cite{Tan240918917}, previous bounds on f-WDM are generally weaker than our limits in the right panel of Figure~\ref{fig:constraints_on_wdm}. We expect our constraints to improve when informed by multiple small-scale structure observables, following recent probe combination studies for $100\%$ WDM models \citep{Enzi201013802,Nadler210107810}. Future data, including dwarf galaxy surveys throughout the Local Volume and population analyses of strong gravitational lenses, will further improve sensitivity to f-WDM \citep{Keeley230107265,Nadler240110318}.

\subsection{Caveats and Future Work}

We now discuss caveats associated with our work. The key caveats detailed in \citetalias{Nadler241003635} are applicable here, including ($i$) the use of DM-only $N$-body simulations inherently does not capture the impact of baryons on (sub)halo populations, ($ii$) we use \textsc{Rockstar} and \textsc{consistent-trees} to identify and link subhalos, which may spuriously lose track of certain subhalos \citealt{Mansfield230810926}), and ($iii$) beyond-CDM physics is assumed to only impact the linear density and velocity transfer functions (thus, for example, we neglect thermal velocities in f-WDM models). As discussed in \citetalias{Nadler241003635}, we plan to address several the first two caveats in future work by resimulating f-NCDM models with embedded galaxy potentials (e.g., following EDEN; \citealt{Wang240801487}), and applying particle tracking-based subhalo finders to our simulations (e.g., following \textsc{Symfind}; \citealt{Mansfield230810926}). Note that the limits presented in this study are robust to these uncertainties because we marginalize over disk disruption in our galaxy--halo connection model and conservatively cut on present-day subhalo mass when deriving our SHMF models. Meanwhile, based on previous simulations, we only expect thermal velocities in the f-WDM models we simulate to impact the very inner regions of (sub)halos, which are not relevant for our SHMF modeling (e.g., \citealt{Maccio12021282}).

There are two additional caveats specific to our results. First, we simulate only one MW-like system. Our constraints on f-NCDM SHMF suppression are thus partly limited by the statistical uncertainties on our simulation measurements. Future work that simulates a larger sample of hosts in f-NCDM models would reduce these uncertainties. Second, this work does not capture potential host-to-host variance in f-NCDM SHMF suppression. However, based on \citetalias{Nadler241003635}, we expect the SHMF suppression to be universal in our f-NCDM models because it is determined strictly by $P(k)$. Thus, our SHMF suppression fits and subsequent results, including our f-NCDM and f-WDM constraints, are not biased by using a single host to derive the SHMF suppression.

Second, while we model f-NCDM SHMF suppression separately for each value of $\delta$, it is in principle possible to model the suppression as a smooth function of $\delta$. This may require a larger suite and/or higher-resolution simulations, which would help ensure that models where SHMF suppression is well resolved do not dominate the fit. Combined with this effort, our simulations will be useful for calibrating semi-analytic predictions for f-NCDM (sub)halo populations, which previously had little direct input from cosmological zoom-in simulations of f-NCDM scenarios. An extended Press--Schechter approach would also provide insight into the scaling of f-NCDM SHMF suppression with $\delta$ and $M_{\mathrm{hm}}$ assumed in our model (Equation~\ref{eq:f_NCDM}). 

\section{Conclusions}
\label{sec:conclusions}

As the second installment of the COZMIC simulation suite, we have presented $24$ cosmological DM-only zoom-in simulations of one MW-like host in f-NCDM scenarios that suppress the linear matter power spectrum, with a plateau in the transfer function on small scales (see Figure~\ref{fig:transfers}). Our work builds on \citetalias{Nadler241003635}, in which $100\%$ beyond-CDM simulations of three MW zoom-in hosts were presented; in another companion study, we simulate models that both suppress $P(k)$ and feature strong, velocity-dependent DM self-interactions (Paper III; \citealt{Nadler241213065}).

Our key results are as follows:
\begin{enumerate}
    \item The SHMF is significantly suppressed in f-NCDM scenarios with transfer function plateau heights of $\delta=0.2$ and $0.4$, for several suppression scales $k_{\mathrm{hm}}$ (Figure~\ref{fig:shmf}).
    \item We derive models for the f-NCDM SHMF suppression that are well constrained for $\delta=0.2$ and $0.4$ (Figure~\ref{fig:shmf_pred}), with best-fit parameters for Equation~\ref{eq:f_NCDM} given by Equations~\ref{eq:best_fit_delta_02} and \ref{eq:best_fit_delta_04}, respectively.
    \item We implement f-NCDM SHMFs in a MW satellite population model to derive $95\%$ confidence lower limits of $k_{\mathrm{hm}}>46$ and $40~\mathrm{Mpc}^{-1}$ for $\delta=0.2$ and $0.4$, respectively (Figure~\ref{fig:constraints_on_khm}); the current data do not constrain $\delta > 0.4$.
    \item We derive the f-WDM limits in Equation~\ref{eq:fwdm_limits}; the current data do not constrain $f_{\mathrm{WDM}}<0.5$ (Figure~\ref{fig:constraints_on_wdm}).
\end{enumerate}

This work, and our choice to simulate a generic f-NCDM transfer function rather than a specific physical model of a fractional beyond-CDM species, is motivated by the idea that small-scale structure observations can probe $P(k)$, even in the scenarios where power is not entirely suppressed on small scales. The results we report, and particularly the overlap between ruled-out f-NCDM transfer functions in the right panel of Figure~\ref{fig:constraints_on_khm}, confirm this expectation. In future work, we plan to generalize these findings to reconstruct $P(k)$ in a model-independent fashion. A larger simulation grid, spanning both f-NCDM models with both different suppression shapes and models that enhance $P(k)$, will be needed to achieve this goal.

\section*{Acknowledgements}

Halo catalogs, merger trees, and particle snapshots are distributed on Zenodo.\footnote{ 10.5281/zenodo.14663119} The analysis code is available on Github.\footnote{\url{https://github.com/eonadler/COZMIC/}}

We are grateful to Wendy Crumrine for comments on the manuscript and to Ariane Dekker, Alex Drlica-Wagner, Subhajit Ghosh, and Chin Yi Tan for helpful discussions. V.G.\ acknowledges the support from NASA through the Astrophysics Theory Program, Award No.\ 21-ATP21-0135, the National Science Foundation (NSF) CAREER grant No. PHY2239205, and from the Research Corporation for Science Advancement under the Cottrell Scholar Program. This research was supported in part by grant NSF PHY-2309135 to the Kavli Institute for Theoretical Physics (KITP).

The computations presented here were conducted through Carnegie's partnership in the Resnick High Performance Computing Center, a facility supported by Resnick Sustainability Institute at the California Institute of Technology. This
work used data from the Milky Way-est suite of simulations\footnote{\url{https://web.stanford.edu/group/gfc/gfcsims/}}, which was supported by the Kavli Institute for Particle Astrophysics and Cosmology at Stanford University, SLAC National Accelerator Laboratory, and the US Department of Energy under contract No.\ DE-AC02-76SF00515 to SLAC National Accelerator Laboratory.

\software{
%Corner \citep{corner},
\textsc{ChainConsumer} \citep{ChainConsumer},
{\sc consistent-trees} \citep{Behroozi11104370},
\textsc{emcee} \citep{emcee},
\textsc{h5py}\footnote{\http{www.h5py.org}},
\textsc{Healpy}\footnote{\http{healpy.readthedocs.io}},
\textsc{Helpers}\footnote{\http{bitbucket.org/yymao/helpers/src/master/}},
\textsc{Pandas} \citep{pandas}, 
% IPython \citep{ipython},
% getdist \citep{getdist},
\textsc{Jupyter}\footnote{\http{jupyter.org}},
\textsc{Matplotlib} \citep{matplotlib},
\textsc{NumPy} \citep{numpy},
\textsc{pynbody} \citep{pynbody},
{\sc Rockstar} \citep{Behroozi11104372}
\textsc{Scikit-Learn} \citep{scikit-learn},
\textsc{SciPy} \citep{scipy},
\textsc{Seaborn}\footnote{\https{seaborn.pydata.org}}.
}

\bibliographystyle{yahapj2}
\bibliography{references,software}

\appendix

\section{Additional Simulation Results}
\label{sec:additional}

\subsection{Host Halo Mass Accretion Histories}
\label{sec:mah}

\begin{figure*}[t!]
\centering
    \includegraphics[trim={0cm 0.3cm 0 -0.75cm},width=\textwidth]{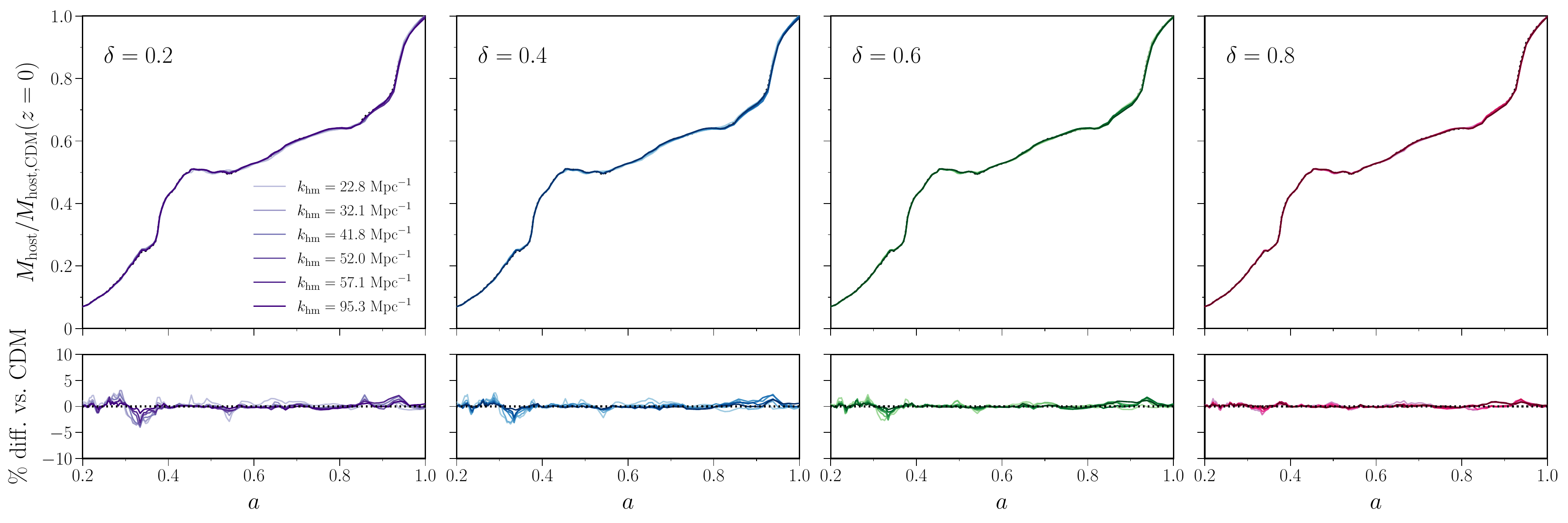}
    \caption{Host halo mass accretion histories for our MW-like host. Black dotted lines (colored lines) show CDM (fractional beyond-CDM) mass accretion histories, with $\delta=0.2$ (first column), $0.4$ (second column), $0.6$ (third column), and $0.8$ (fourth column). All mass accretion histories are normalized to the CDM host's mass at $z=0$. Bottom panels show the percent difference between each fractional beyond-CDM model and CDM. Host masses in our fractional beyond-CDM simulations match CDM at the percent level at all redshifts.}
    \label{fig:mah}
\end{figure*}

Figure~\ref{fig:mah} shows our host's mass accretion history, normalized to its virial mass in CDM at $z=0$, for all of our f-NCDM simulations. Our MW-like host undergoes an early merger with a Gaia Sausage-Enceladus (GSE) analog system, and accretes a massive LMC analog subhalo at late times. In our f-NCDM simulations, the host's mass accretion history matches CDM at the percent level at all redshifts. This similarity is expected given that $P(k)$ is nearly identical to CDM on the scales corresponding to our MW host in all f-NCDM models we simulate.

\subsection{Subhalo Radial Distributions}
\label{sec:basic_rad}

\begin{figure*}[t!]
\centering
\includegraphics[width=\textwidth]{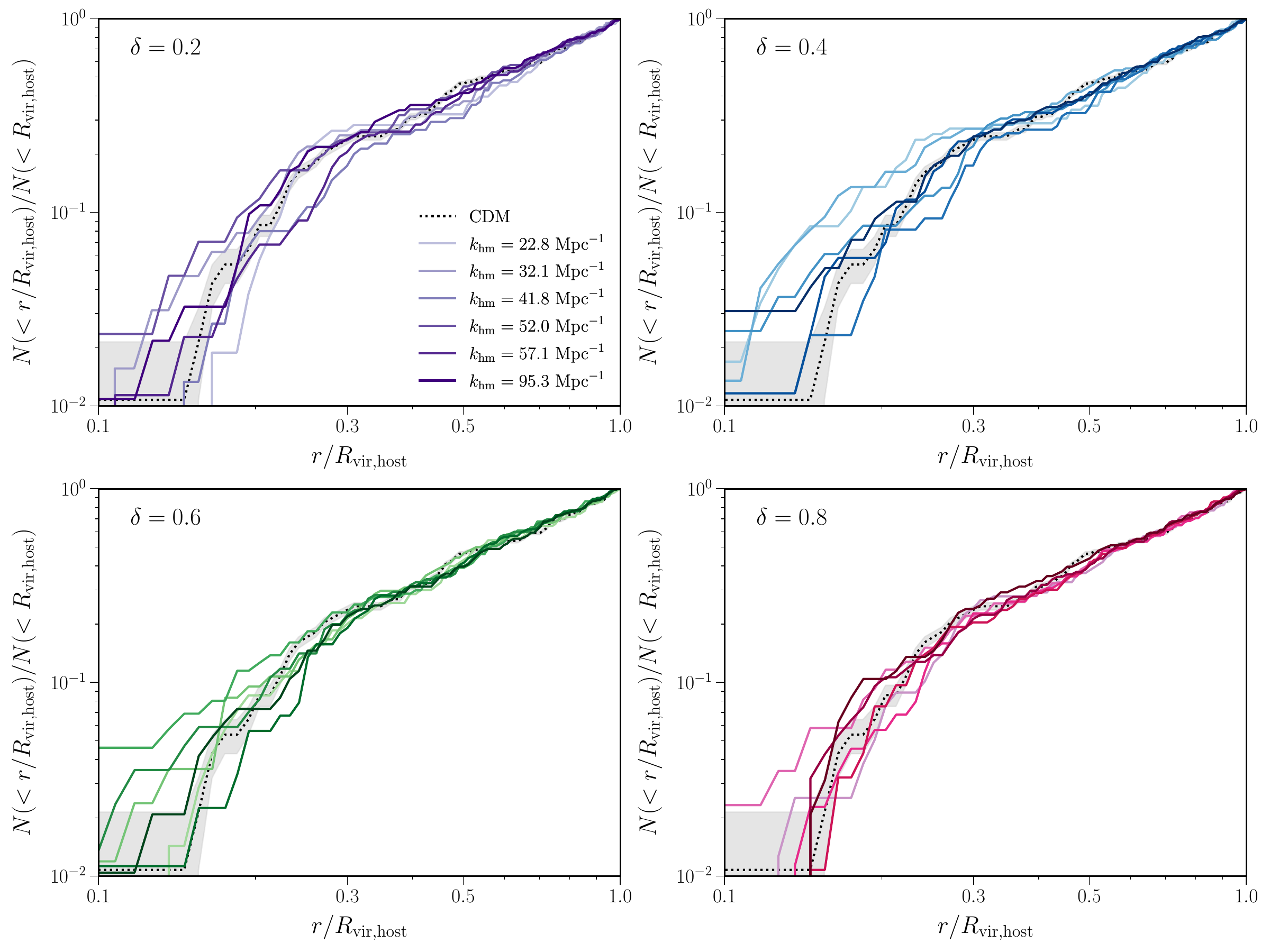}
    \caption{Subhalo radial distributions for the MW-like host in f-NCDM (solid colored lines) and in CDM (dotted black lines). The cumulative number of subhalos as a function of distance from the host center in units of the virial radius $r/R_{\mathrm{vir,host}}$, is shown for $\delta = 0.2$ (top left), $0.4$ (top right), $0.6$ (bottom left), and $0.8$ (bottom right); suppression scales range from $k_{\mathrm{hm}}=22.8$ (lightest) to $95.3~\mathrm{Mpc}^{-1}$ (darkest) in each panel. Each radial distribution is normalized to the total number of subhalos within $R_{\mathrm{vir,host}}$. All results are restricted to subhalos with present-day virial masses $M_{\mathrm{sub}}>1.2\times 10^8~M_{\mathrm{\odot}}$. Gray bands show the $1\sigma$ Poisson uncertainty on the CDM radial distribution.}
    \label{fig:radial}
\end{figure*}

Figure~\ref{fig:radial} shows the normalized radial distribution of subhalos in the MW-like host for each f-NCDM simulation. We normalize radial distributions to the total number of subhalos within the virial radius to isolate trends in their shape, since total subhalo abundances differ among our models. We apply the mass cut $M_{\mathrm{sub}}>1.2\times 10^8~M_{\mathrm{\odot}}$ in all cases. Note that the virial radius for Halo004 is $R_{\mathrm{vir,host}}\approx 300~\mathrm{kpc}$ in all simulations. The normalized f-NCDM radial distributions are scattered about the CDM result. This scatter is larger than the Poisson uncertainty on the CDM radial distribution and is comparable to the host-to-host variance measured across three MW zoom-in simulations in \citetalias{Nadler241003635}. Note that this scatter is partly due to stochasticity in subhalos' orbital evolution when ICs are varied.

Nonetheless, Figure~\ref{fig:radial} indicates that f-NCDM models with more severe $P(k)$ suppression may yield more concentrated radial distributions. This is consistent with \cite{Lovell210403322}, where this effect was interpreted as a result of the suppression of low-mass subhalos in corresponding models. Because lower-mass subhalos tend to be less radially concentrated than subhalos with large infall masses, the subhalo population that survives in f-NCDM models is slightly more concentrated. However, due to the stochasticity discussed above, this effect is not visible in every simulation; for example, compare the $\delta=0.2$ and $\delta=0.4$ panels in Figure~\ref{fig:radial}. 

In future work, dedicated study of subhalo orbital properties and tidal evolution using a larger sample of f-NCDM simulations, combined with semi-analytic models, will be useful to isolate the physical differences between f-NCDM and CDM radial distributions. When deriving MW satellite limits in Section~\ref{sec:limits}, we assume that the radial distribution is unchanged in f-NCDM scenarios relative to CDM, and thus that the differences hinted at in Figure~\ref{fig:radial} are statistical fluctuations rather than systematic effects.

\subsection{SHMF Suppression for $\delta=0.6$ and $0.8$}
\label{sec:other_delta}

In Section~\ref{sec:shmf}, we presented SHMF fits for $\delta=0.2$ and $0.4$ because these scenarios are constrained by our MW satellite inference. Here, we present the $\delta=0.6$ and $0.8$ results.

For $\delta=0.6$, we obtain $\alpha=2.1^{+17.0}_{-1.9}$, $\beta=0.8^{+1.3}_{-0.7}$, and $\gamma=0.2^{+0.5}_{-0.2}$, at $68\%$ confidence. The parameters that maximize our posterior for $\delta=0.6$ are
\begin{align}
    \alpha_{\delta=0.6}=3.8,\nonumber\\
    \beta_{\delta=0.6}=1.1,\nonumber\\
 \gamma_{\delta=0.6}=0.6.\label{eq:best_fit_delta_06}
\end{align}

For $\delta=0.8$, we obtain $\alpha=1.1^{+31.2}_{-1.0}$, $\beta=0.0^{+0.9}_{-0.0}$, and $\gamma=0.1^{+0.7}_{-0.1}$, at $68\%$ confidence. The parameters that maximize our posterior for $\delta=0.8$ are
\begin{align}
    \alpha_{\delta=0.8}=41.2,\nonumber\\
    \beta_{\delta=0.8}=0.3,\nonumber\\
 \gamma_{\delta=0.8}=0.2.\label{eq:best_fit_delta_08}
\end{align}

\begin{figure}[t!]
\centering
\hspace{-5mm}
\includegraphics[trim={0 0.35cm 0 0cm},width=0.49\textwidth]{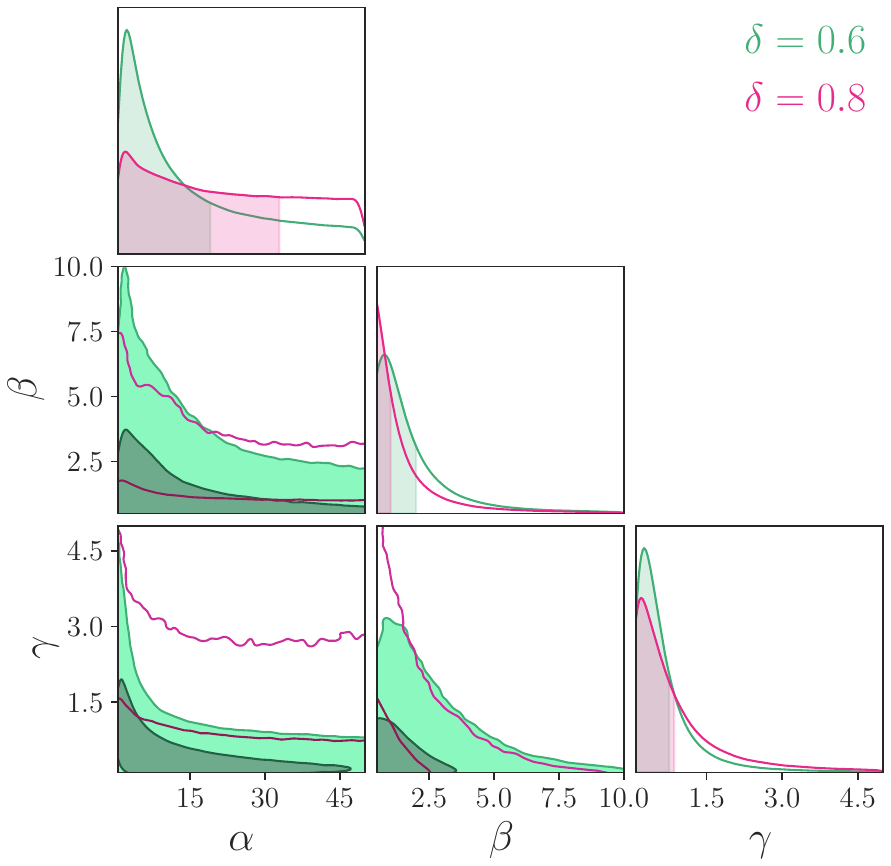}
    \caption{Marginalized posteriors from our f-NCDM SHMF suppression fit for $\delta=0.6$ (green) and $0.8$ (pink). Dark (light) two-dimensional contours show $68\%$ ($95\%$) confidence intervals; top and side panels show marginal posteriors with shaded $68\%$ confidence intervals.}
    \label{fig:shmf_posterior_alt}
\end{figure}

Figure~\ref{fig:shmf_posterior_alt} compares the posteriors from our $\delta=0.6$ and $0.8$ SHMF fits. Comparing these posteriors to each other and to Figure~\ref{fig:shmf_posterior}, we conclude that SHMF becomes less well constrained as $\delta$ increases. In the case of $\delta=0.8$, uncertainties on all parameters are particularly large and the inferred SHMF suppression is nearly independent of $M_{\mathrm{sub,peak}}$. Due to these large uncertainties, the SHMF suppression parameters in both cases are consistent with those inferred in our $\delta=0.2$ and $0.4$ scenarios. We expect that even higher-resolution simulations will be needed to improve SHMF suppression constraints for $\delta=0.6$ and $0.8$.

\begin{figure*}[t!]
\includegraphics[width=0.5\textwidth]{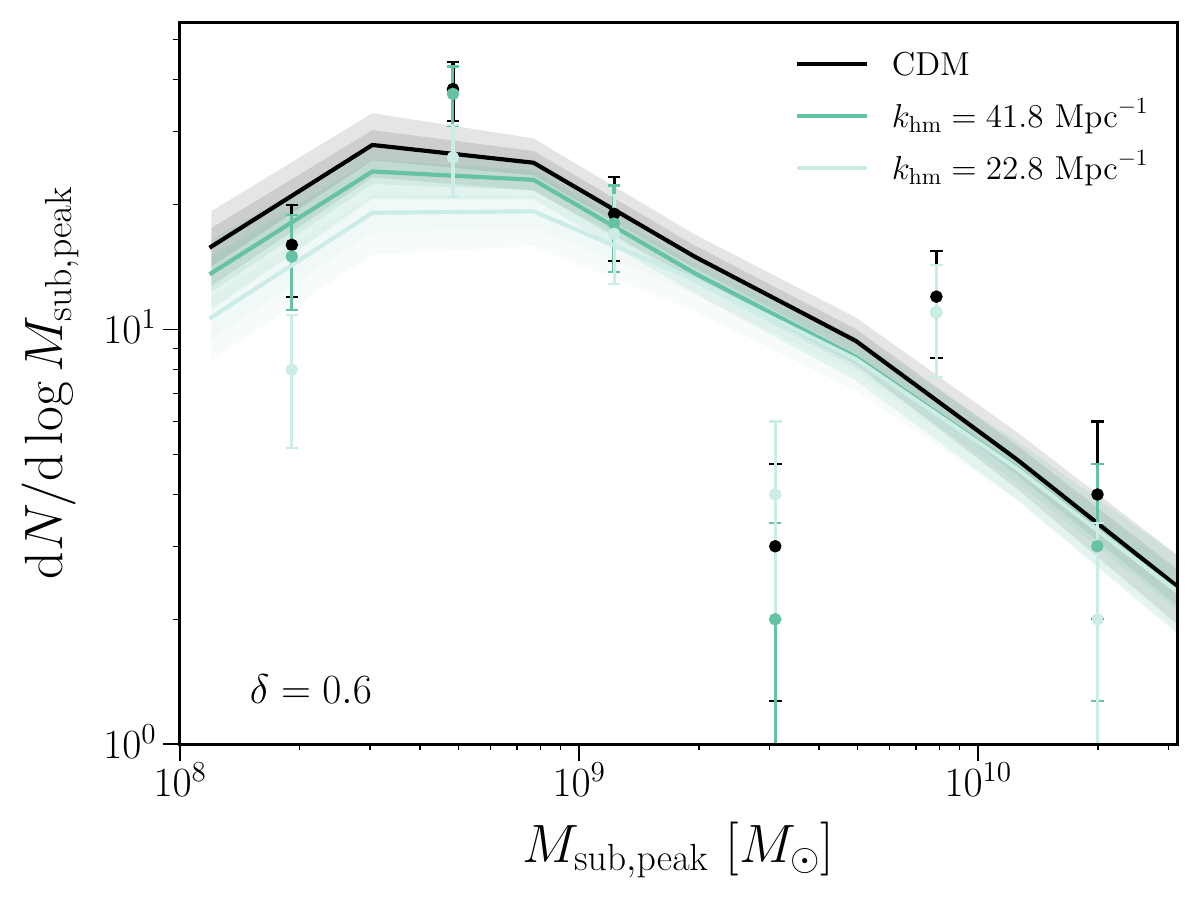}
\includegraphics[width=0.5\textwidth]{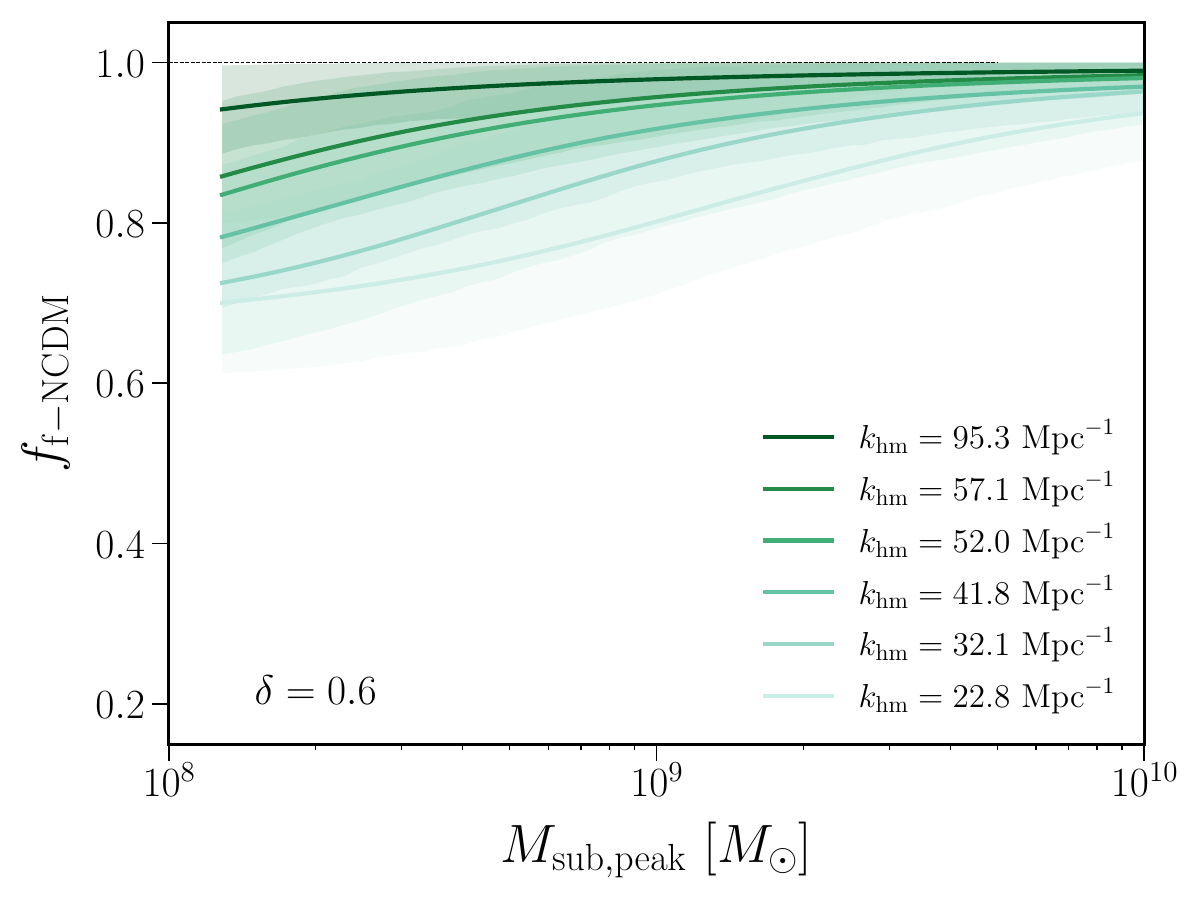}\\
\includegraphics[width=0.5\textwidth]{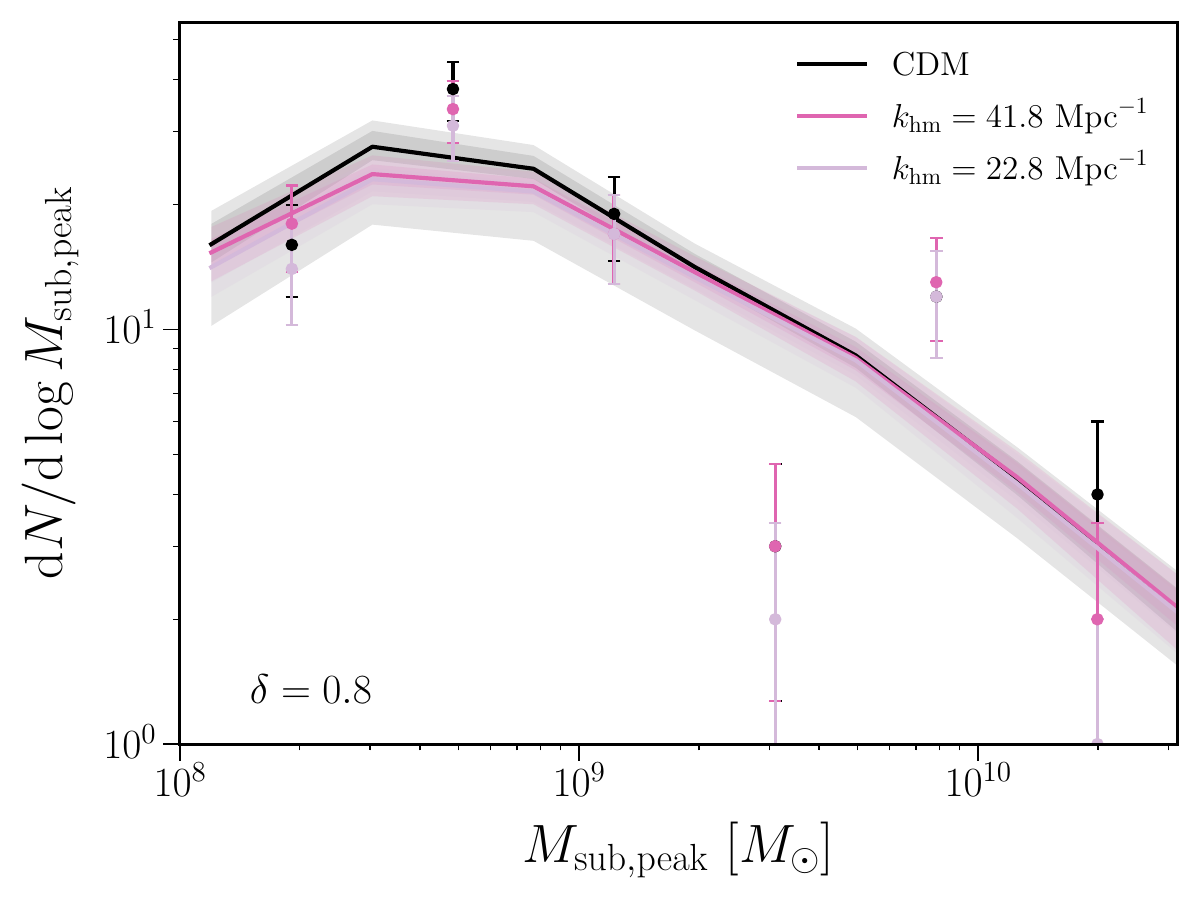}
\includegraphics[width=0.5\textwidth]{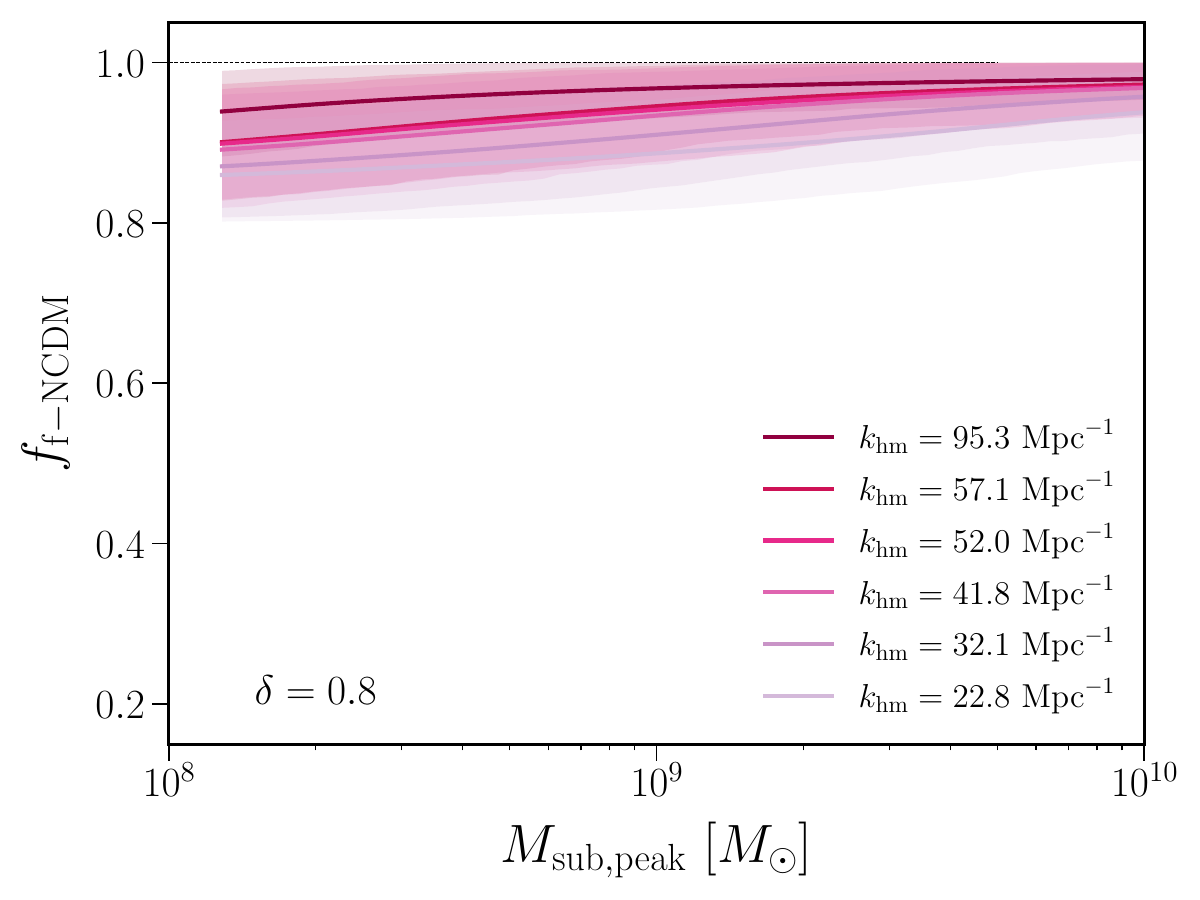}
    \caption{\emph{Left}: Mean differential SHMF for our MW-like host, as a function of peak virial mass, for f-NCDM simulations with $\delta=0.6$ (top; green), $\delta=0.8$ (bottom; pink), and CDM (black). Error bars show the $1\sigma$ Poisson error on the mean differential SHMF. \emph{Right}: Corresponding SHMF suppression predictions. In all panels, dark (light) bands show $68\%$ ($95\%$) confidence intervals from our SHMF fit, calculated by sampling from the posterior.}
    \label{fig:shmf_pred_alt}
\end{figure*}

Figure~\ref{fig:shmf_pred_alt} shows the differential SHMF and its suppression in $\delta=0.6$ (top) and $0.8$ (bottom). Our $\delta=0.6$ model fits the simulation measurements fairly well ($p=0.16$, two-sample KS test; $\chi^2=0.6$). A meaningful fit is possible because the SHMF is significantly suppressed relative to CDM for the most suppressed model we simulate in this scenario ($k_{\mathrm{hm}}=22.8~\mathrm{Mpc}^{-1}$; see Figure~\ref{fig:shmf}). Meanwhile, no models in the $\delta=0.8$ scenario significantly suppress the SHMF, so the fit is not constraining, although the goodness of fit is similar ($p=0.1$, two-sample KS test; $\chi^2=0.6$). Thus, the suppression in the bottom-right panel of Figure~\ref{fig:shmf_pred_alt} largely reflects the SHMF suppression plateau assumed in our model (Equation~\ref{eq:f_NCDM}).

\section{Convergence Tests}
\label{sec:convergence}

We perform four additional HR simulations with $k_{\mathrm{hm}}=22.8~\mathrm{Mpc}^{-1}$ ($m_{\mathrm{WDM}}=3~\mathrm{keV}$) and $\delta\in [0.2~, 0.4,~ 0.6,~ 0.8]$. These HR simulations use five refinement regions, yielding the highest-resolution region with $16,384$ particles per side and $m_{\mathrm{part,high-res}}=5.0\times 10^4~M_{\mathrm{\odot}}$. We set the gravitational softening to $\epsilon=80~\mathrm{pc}~h^{-1}$.

\begin{figure}[t!]
\hspace{-5mm}
\includegraphics[width=0.49\textwidth]{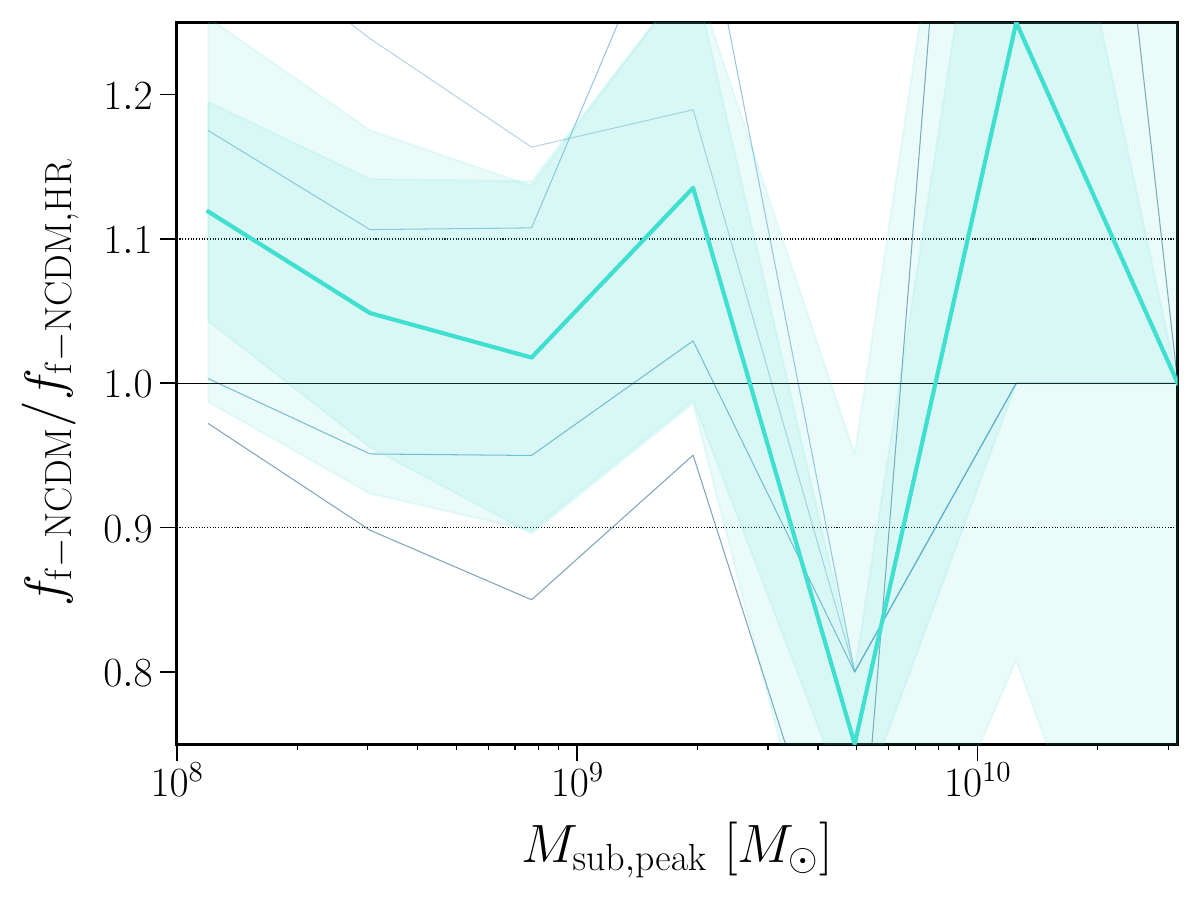}
    \caption{Ratio of the differential f-NCDM peak SHMF relative to CDM, for subhalos with $M_{\mathrm{sub}}>1.2\times 10^8~M_{\mathrm{\odot}}$ in our fiducial-resolution simulations of Halo004, relative to the same quantity in our HR resimulations of the same host. The bold line shows the mean ratio stacked across all five f-NCDM models with $\delta\in [0.2,~ 0.4,~ 0.6,~ 0.8]$ and $k_{\mathrm{hm}}=22.8~\mathrm{Mpc}^{-1}$. Dark bands show $1\sigma$ Poisson uncertainty on the mean ratio, and light bands show model-to-model scatter.}
    \label{fig:shmf_convergence}
\end{figure}

Figure~\ref{fig:shmf_convergence} shows the results of this convergence test by plotting the ratio of the SHMF suppression in our standard-resolution f-NCDM simulations to that in our HR runs. For the HR measurement, we normalize to the HR CDM SHMF from \citetalias{Nadler241003635} (originally presented in \citealt{Buch240408043}). Our f-NCDM measurements are converged at the $\approx 10$--$20\%$ level for all $M_{\mathrm{sub,peak}}$ we consider. The uncertainty on the fiducial-to-high resolution SHMF suppression ratio is dominated by Poisson error rather than model-to-model scatter because we simulate only one MW-like host. As demonstrated in \citetalias{Nadler241003635}, the statistical uncertainty is reduced by simulating more hosts, but the model-to-model scatter persists.

\end{document}